\documentclass[twocolumn,prl, aps,superscriptaddress]{revtex4}

\usepackage{mathptmx}
\usepackage{sublabel}
\usepackage{dcolumn}
\usepackage{subfigure}
\usepackage{amsmath}
\usepackage{amssymb}
\usepackage{amsfonts}
\usepackage{bm}
\usepackage{bbm}
\usepackage{color}
\usepackage{overpic}
\usepackage{latexsym}
\usepackage{epstopdf}
\usepackage{color}
\usepackage[english]{babel}
\usepackage{latexsym}
\usepackage{stmaryrd}

\usepackage{psfrag,graphicx}
\usepackage{epsf}
\usepackage{natbib}
\usepackage{epstopdf}
\usepackage{appendix}
\usepackage{dsfont}
\usepackage{mathtools}

\definecolor{mygrey}{gray}{0.35}
\definecolor{myblue}{rgb}{0.2,0.2,0.8}
\definecolor{myzard}{cmyk}{0,0,0.05,0}
\definecolor{mywhite}{rgb}{1,1,1}
\definecolor{myred}{rgb}{1,0.,0.3}

\usepackage[colorlinks=true,citecolor=myblue,linkcolor=myred]{hyperref}
\def\beq{\begin{equation}}
\def\eeq{\end{equation}}
\def\ba{\begin{align}}
\def\enda{\end{align}}
\def\bi{\begin{itemize}}
\def\ei{\end{itemize}}

\newcommand{\eq}[1]{Eq.\;(\ref{#1})}

 \def\ee{\mathord{\rm e}}
 
 \def\ii{\mathord{\rm i}}

\def\tr{\mathop{\rm Tr}}

 \def\ee{\mathord{\rm e}}
 
 \def\ii{\mathord{\rm i}}

\def\tr{\mathop{\rm Tr}}

\renewcommand{\ii}{{\rm i}}
\renewcommand{\ee}{{\rm e}}

 \newcommand{\ket}[1]{|#1\rangle}
 \newcommand{\bra}[1]{\langle #1|}

 \newcommand{\ketbradif}[2]{\ket{#1}\bra{#2}}
 \newcommand{\ketbra}[1]{\ketbradif {#1}{#1}}

\graphicspath{{Plots/}}

\begin{document}

\title[Short Title]{Two-dimensional spectroscopy for the study of ion Coulomb crystals}

\author{A. Lemmer}
\affiliation{Institut f\"ur Theoretische Physik, Albert-Einstein Alle 11,
Universit\"at Ulm, 89069 Ulm, Germany}

\author{C. Cormick}
\affiliation{Institut f\"ur Theoretische Physik, Albert-Einstein Alle 11,
Universit\"at Ulm, 89069 Ulm, Germany}

\author{C. Schmiegelow}
\affiliation{QUANTUM, Institut f\"ur Physik, Universit\"at Mainz, 55128 Mainz, Germany}

\author{F. Schmidt-Kaler}
\affiliation{QUANTUM, Institut f\"ur Physik, Universit\"at Mainz, 55128 Mainz, Germany}

\author{M. B. Plenio}
\affiliation{Institut f\"ur Theoretische Physik, Albert-Einstein Allee 11,
Universit\"at Ulm, 89069 Ulm, Germany}

\begin{abstract}
Ion Coulomb crystals are currently establishing themselves as a highly controllable
test-bed for mesoscopic systems of statistical mechanics. The detailed experimental
interrogation of the dynamics of these crystals however remains an experimental
challenge. In this work, we show how to extend the concepts of multi-dimensional 
nonlinear spectroscopy to the study of the dynamics of ion Coulomb crystals.
The scheme we present can be realized with state-of-the-art technology and gives direct access to the dynamics,
revealing nonlinear couplings even in the presence of thermal excitations. 
We illustrate the advantages of our proposal showing how two-dimensional
spectroscopy can be used to detect signatures of a structural phase transition of the ion
crystal, as well as resonant energy exchange between modes. Furthermore, we demonstrate
in these examples how different decoherence mechanisms can be identified.
\end{abstract}


\date{\today}
\maketitle

Two-dimensional (2D) spectroscopy was first proposed and realized in the context of nuclear magnetic 
resonance (NMR) experiments and has proven to be a very valuable tool in the investigation of complex 
spin systems \cite{ernst_buch}. By properly designed pulse sequences complicated spectra can be unravelled by the separation
of interactions originating from different physical mechanisms to different frequency axes. The method allows for the estimation of spin-spin couplings
in complex spin systems and the identification of different sources of noise. 2D spectroscopy has been
adapted with remarkable success to other fields, facilitating the investigation of anharmonic molecular vibrational spectra in
the infrared \cite{2D_IR_book}, electronic dynamics in molecular aggregates \cite{Mukamel_book}
and photosynthetic pigment-protein complexes \cite{2D_photosynthesis_experiments}, and photochemical reactions~\cite{2d_photochemistry}.

Here we propose and analyze the application of 2D spectroscopy for the precise experimental characterization of nonlinear
dynamics in few-or many-body systems of interest for quantum optics, in particular, in trapped-ion Coulomb crystals. The
excellent control over the internal and motional degrees of freedom makes trapped atomic ions~\cite{trapped_ions_review_wineland} a versatile tool to study
statistical mechanics of systems in and out of equilibrium \cite{spin_simulation_ions, Kibble-Zurek_ions, heat_transport_ions}.
A paradigmatic example is provided by the linear-to-zigzag structural transition \cite{linear-zigzag_transition, quantum_linear-zigzag_transition}.
In the vicinity of the transition, the usual harmonic treatment of the motion breaks down and nonlinear terms in the potential are essential
for understanding the dynamics of the Coulomb crystal. Nonlinearities added to the
trap potential have also been proposed for the implementation of the Frenkel-Kontorova model \cite{Frenkel-Kontorova} and the Bose-Hubbard model \cite{BHM_Porras}.
The scheme we present can be used for the analysis of
nonlinear dynamics, and, more generally, it represents a new appproach for the interrogation of complex quantum systems constructed from ion
crystals. Some features of 2D spectroscopy are especially appealing in this context: it can provide information that is not
accessible in 1D Ramsey-type experiments, it can filter out the contribution from purely harmonic terms, and it
allows to distinguish dephasing and relaxation due to environmental dynamical degrees of freedom from fluctuations between
subsequent experimental runs. We note that, as opposed to a related scheme \cite{Gessner-arXiv}, our
proposal requires neither the technically demanding individual addressing of ions in the Coulomb crystal nor ground-state cooling.
Furthermore, a purely harmonic evolution produces no 2D spectroscopic signal in our protocol~\cite{sup_mat_f}. We expect that these properties
constitute key elements for the investigation of nonlinear dynamics in large crystals \cite{Drewsen, Bollinger}. After a brief review of
the general formalism of 2D spectroscopy we illustrate its usefulness in ion-trap experiments with two case examples.

{\it 2D spectroscopy} \cite{2D_IR_book,ernst_buch,Mukamel_book}. After state initialization, a general multidimensional spectroscopy
experiment consists of a sequence of $n$ electromagnetic pulses on the system under investigation
separated by intervals of free evolution. The action of the $k$th pulse on the
system's density matrix is described by a superoperator $\hat{P}_k$. It is followed by a period of time $t_k$ in which
the system evolves under a Hamiltonian $H_k$, with an associated superoperator $\hat{H}_k$, and additional dissipative processes
described by $\hat{\Gamma}_k$ resulting in a Lindblad superoperator $\hat{L}_k = -\ii \hat{H}_k - \hat{\Gamma}_k$. The temporal
variables $t_k$ are scanned over an interval $[0, t_{k}^{\rm max}]$ and at the end of every experiment an operator $M$ is measured
giving a signal:
\begin{eqnarray}
&&s(t_1,\ldots,t_n)=\tr [M \rho(t_1,\ldots,t_n)], \label{eq:signal}\\
&&\rho(t_1,\dots,t_n) = \exp[\hat{L}_n t_n] \hat{P}_n \cdots \exp[\hat{L}_1 t_1] \hat{P}_1 \rho_0 \label{eq:rho}
\end{eqnarray}
where $\rho_0$ is the initial state and we assume for simplicity that $\hat{L}_k$ is time independent. The
frequency-domain signal, which contains spectral information of the Liouvillians governing the free evolution
periods, is extracted by a Fourier transform of the signal $s(t_1,\ldots,t_n)$ in one or several time variables.
A two-dimensional spectrum displays the signal as a function of two of the time or frequency variables.

In the implementation we propose, the pulses correspond to phase-controlled displacements $\hat P_k \rho = D(\alpha_k) \rho D(\alpha_k)^\dagger$ on one of the motional modes
of the ion crystal. Here, $D(\alpha_k) = \exp[\alpha_k a^\dagger - \alpha_k^* a]$ with $\alpha_k=|\alpha_k|\ee^{\ii \phi_k}$ and $a$ the annihilation operator of the mode. 
We consider sequences involving four such pulses, followed by a measurement of the mode population.
For small $\alpha_k$, the displacement operators can be expanded in powers of $\alpha_k$.  Using this expansion 
and phase cycling, one can identify the coherence transfer pathways that contribute to the final signal~\cite{ernst_buch,sup_mat_d}.
This allows for an understanding of the physical origin of each spectral peak.

{\it Nonlinear terms in the Coulomb interaction between trapped ions.}
We consider $N$ singly-charged ions of mass $m$ in a linear Paul trap described by an effective harmonic confining potential.
Taking into account the mutual Coulomb repulsion between ions the Hamiltonian of the system reads
\beq
    H= \sum_{i,\mu} \left( \frac{p_{i\mu}^2}{2 m} + \frac{1}{2} m \omega_{\mu}^2 r_{i\mu}^2 \right)
    + \frac{1}{2}\sum_{i \neq j} \frac{e^2}{4\pi \epsilon_0} \frac{1}{|{\bf r}_i - {\bf r}_j|}.
    \label{full_ham}
\eeq
Here $\{ \omega_{\mu} \}_{\mu=x,y,z}$ denote the trap frequencies, $r_{i\mu}\:(p_{i\mu})$ the position
(momentum) of ion $i$ in spatial direction $\mu$, and $\epsilon_0$ the vacuum permittivity. If $\omega_x,\omega_y
\gg \omega_z$, cold ions arrange on a string along the $z$-axis and perform small oscillations 
$\delta r_{i\mu}(t) =r_{i\mu}(t) - r_{i\mu}^0 $ about their equilibrium positions $r_{i\mu}^0$.
The Hamiltonian expanded to second order in $\delta r_{i\mu} (t)$ can be diagonalized so that the motional degrees of
freedom are described by a set of $3N$ uncoupled harmonic oscillators:
\beq
    H \approx H_0 = \hbar \omega_z \sum_{n} \left( \sqrt{\gamma^x_n}  a_n^{\dagger} a_n + \sqrt{\gamma^y_n}
    b_n^{\dagger} b_n + \sqrt{\lambda^z_n} c_n^{\dagger} c_n \right)\,.
    \label{normal_mode_ham}
\eeq
Here, $a_n\:(b_n,c_n)$ denotes the annihilation operator for mode $n$ in $x\:(y,z)$ direction and
$\lambda_n^z$ and $\gamma_{n}^{x/y}$ are the eigenvalues of the Hessian matrices of the potential in the different spatial directions.
In each direction, $n=1$ denotes the center-of-mass mode and $n=N$ the mode where neighboring ions move in counterphase.
In transverse directions this mode is dubbed the zigzag (zz) mode.

We consider a linear chain along $z$ with $\omega_y > \omega_x $ and focus on dynamics involving transverse motion
in $x$-direction. The first, nonlinear, corrections to $H_0$ arise with the third and
fourth-order terms in the Taylor expansion of the Coulomb potential \cite{james_third_order_main}:
\beq
H^{(3)}= 3 \frac{z_0}{4 l_z} \hbar \omega_z \sum_{n,m,p} \frac{D_{nmp}^{(3)}}{\sqrt[4]{\gamma_n^x \gamma_m^x \lambda_p^z}}  (a_n + a_n^{\dagger}) (a_m+a_m^{\dagger}) ( c_p
+c_p^{\dagger}) \,,
\label{third_order_ham_main}
\eeq
and~\cite{sup_mat_c}
\beq
\begin{split}
H^{(4)} = & 3 \left( \frac{z_0}{4 l_z} \right)^2  \hbar \omega_z \sum_{n,m,p,q} D_{nmpq}^{(4)} \frac{(a_n + a_n^{\dagger}) (a_m + a_m^{\dagger})}{\sqrt[4]{\gamma_n^x \gamma_m^x}} \\
 & \times \left[  \frac{(a_p + a_p^{\dagger}) (a_q + a_q^{\dagger}) }{\sqrt[4]{\gamma_p^x \gamma_q^x}}
 +\frac{2 (b_p + b_p^{\dagger}) (b_q + b_q^{\dagger})}{\sqrt[4]{\gamma_p^y \gamma_q^y}}   \right. \\
& \left. -\frac{8 (c_p + c_p^{\dagger}) (c_q + c_q^{\dagger})}{\sqrt[4]{\lambda_p^z \lambda_q^z}}  \right]\,. \\
\end{split}
\label{fourth_order_ham}
\eeq
Here, $z_0=\sqrt{\hbar/(2m\omega_z)}$ is the spread of the ground-state wavefunction for the axial center-of-mass mode
and $l_z=[e^2/(4\pi\epsilon_0 m \omega_z^2)]^{1/3}$ is the length scale of the inter-ion spacing set by the axial trapping~\cite{sup_mat_a}, while $D^{(3)}_{nmp}$
and $D^{(4)}_{nmpq}$ depend on the dimensionless equilibrium positions and normal-mode coefficients. We report only
terms involving modes in $x$-direction \cite{sup_mat_c}.

Under typical operating conditions $z_0/4l_z \approx 10^{-3}$ as $\omega_z$ usually lies in the MHz range.
This implies that third-order contributions of the perturbation expansion represent small corrections to the harmonic Hamiltonian $H_0$.
 However, the trap frequencies can be tuned to resonances so that there is coherent
energy transfer between modes~\cite{james_third_order_main}. In this regime, nonlinear terms
cannot be neglected. We note that such resonances become generic in systems with many ions. Sufficiently
far from resonances, the dominant effect of the third-order terms is given by Kerr-type shifts of the
mode frequencies found in second-order perturbation theory~\cite{Nie-Roos-James-2009_main}. The fourth-order contributions
in the Taylor expansion of the Coulomb potential also result in such shifts. Both contributions are
smaller than the harmonic terms by roughly a factor $(z_0/4l_z)^2\approx 10^{-6}$. These small cross-Kerr nonlinearities
can become important in quantum information experiments where shifts of the order of $1-20\,$Hz were found to affect the 
achieved fidelity~\cite{innsbruck_cross_kerr_main}. Moreover, fourth-order
contributions of the Coulomb potential are fundamental for the description of structural transitions
such as the linear-to-zigzag transition \cite{linear-zigzag_transition, quantum_linear-zigzag_transition}.

In the following we analyze how to access the nonlinear dynamics of the ions by means of 2D spectroscopy. To
this end we consider a linear string of $N=3$ ions, which displays the essential characteristics of nonlinear
mode coupling, while the reduced complexity of the 2D spectra facilitates their interpretation.
With increasing system size the linear spectrum becomes more crowded, resonances may appear without being deliberately
tuned, and ground-state cooling of all modes becomes harder, thus making cross-Kerr energy shifts more problematic. 
As 2D spectroscopy can deal with all of these problems it becomes increasingly useful with increasing system size.

\begin{table*}[hbt]
\caption{Simulation parameters for the 2D spectrum in the neighbourhood of the linear-to-zigzag transition in Fig.~\ref{fig_fourth_order_spectrum}. Definitions are given in the main text.}
\renewcommand{\arraystretch}{1.5}
\begin{tabular}{p{1.2cm}p{1.7cm}p{1.2cm}p{1.7cm}p{1.4cm}p{1.4cm}p{1.4cm}p{1.4cm}p{1.4cm}p{1.4cm}p{1cm}p{1cm}}
\hline
\hline
$\omega_z/2\pi\,$ & $\omega_x/2\pi$ & $\omega_y/2\pi$ & $\omega_{\rm zz}/2\pi$ & $t_{\rm 1/3}^{\rm max}$ & $\Delta t_{1/3}$ & $ \Delta \omega_{\rm
zz}/2\pi$
& $\Omega_{\rm SI}/2\pi$ & $\Omega_{\rm d,3}^y/2\pi$ & $\Omega_{\rm d,3}^z/2\pi$ &$|\alpha_{k}|$ & $N_{\phi_k}$ \\
\hline
2$\,$MHz &  3.1012$\,$MHz & 5$\,$MHz & 131.95$\,$kHz & 2$\,$ms & 25.3$\,\mu$s & 15.20$\,$kHz & 5.12$\,$kHz & 0.58$\,$kHz &
-1.37$\,$kHz & 0.25 & 4 \\
\hline
\hline
\end{tabular}
\label{tab_fourth_order_parameters}
\end{table*}

{\it Signatures of the onset of a structural transition from 2D spectroscopy.}
The linear-to-zigzag transition occurs when the confining potential in one radial direction is reduced below a critical value
at which the ions break out of the linear structure. We consider a case in which the potential in $x$-direction is lowered approaching, 
but not crossing, the linear-to-zigzag transition. On approach to the structural transition, the zz-mode frequency 
$\omega_{\rm zz}=\sqrt{\gamma_{\rm zz}^x}\omega_z$ approaches zero as $\gamma_{\rm zz}^x $ goes to zero. 
This leads to an increase of the fourth-order terms 
in~\eq{fourth_order_ham} involving $\gamma_{\rm zz}^x$. The increase is fastest for the term
whose coefficient scales as~$1/\gamma_{\rm zz}^x$ which contains non-rotating terms $\propto (a_{\rm zz}^{\dagger})^2 a_{\rm zz}^2$ 
and $ \propto a_{\rm zz}^{\dagger} a_{\rm zz}$. The former corresponds to a self-interaction of the zz-mode that introduces an 
energy penalty when placing more than one phonon in the mode, while the second term shifts the zz-mode frequency.

The effects of the third-order Hamiltonian~\eq{third_order_ham_main} are comparable to
the contributions of the fourth-order terms, but carry opposite signs so partial cancelations occur. In an interaction picture
with respect to the normal modes, we obtain an effective Hamiltonian $H_{\rm eff}^{(4)} = H_{\rm s} + H_{\rm d}$
consisting of the self-interaction (SI) part
\beq
    H_{\rm s} = \hbar\frac{\Omega_{\rm SI}}{2} (a^{\dagger}_{\rm zz})^2 a_{\rm zz}^2 + \hbar\Delta \omega_{\rm zz} a_{\rm zz}^{\dagger} a_{\rm
    zz}\,,
    \label{fourth_order_shift_ham}
\eeq
and a dephasing part arising from cross-Kerr couplings:
\beq
    H_{\rm d} = a_{\rm zz}^{\dagger}a_{\rm zz} \left(\hbar \Omega_{\rm d,2}^x  a^{\dagger}_2 a_2
    +  \hbar \sum_{n=2,3} \Omega_{\rm d,n}^{y}  b_n^{\dagger} b_n + \Omega_{\rm d,n}^{z} c_n^{\dagger} c_n\right).
    \label{fourth_order_dephasing_ham}
\eeq
The self-interaction strength is $\Omega_{\rm SI} = 36 (z_0/4l_z)^2 \omega_z D^{(4)}_{3333}/\gamma_{\rm zz}^x$
while the dephasing rates $\Omega_{\rm d,n}^{\mu}$ scale as $1/ \sqrt{\gamma_{\rm zz}^x}$. The correction $\Delta
\omega_{\rm zz}$ to the zz-mode frequency is mainly due to self-interaction~\cite{sup_mat_c}.

\begin{figure}[b]
\includegraphics[width=\columnwidth]{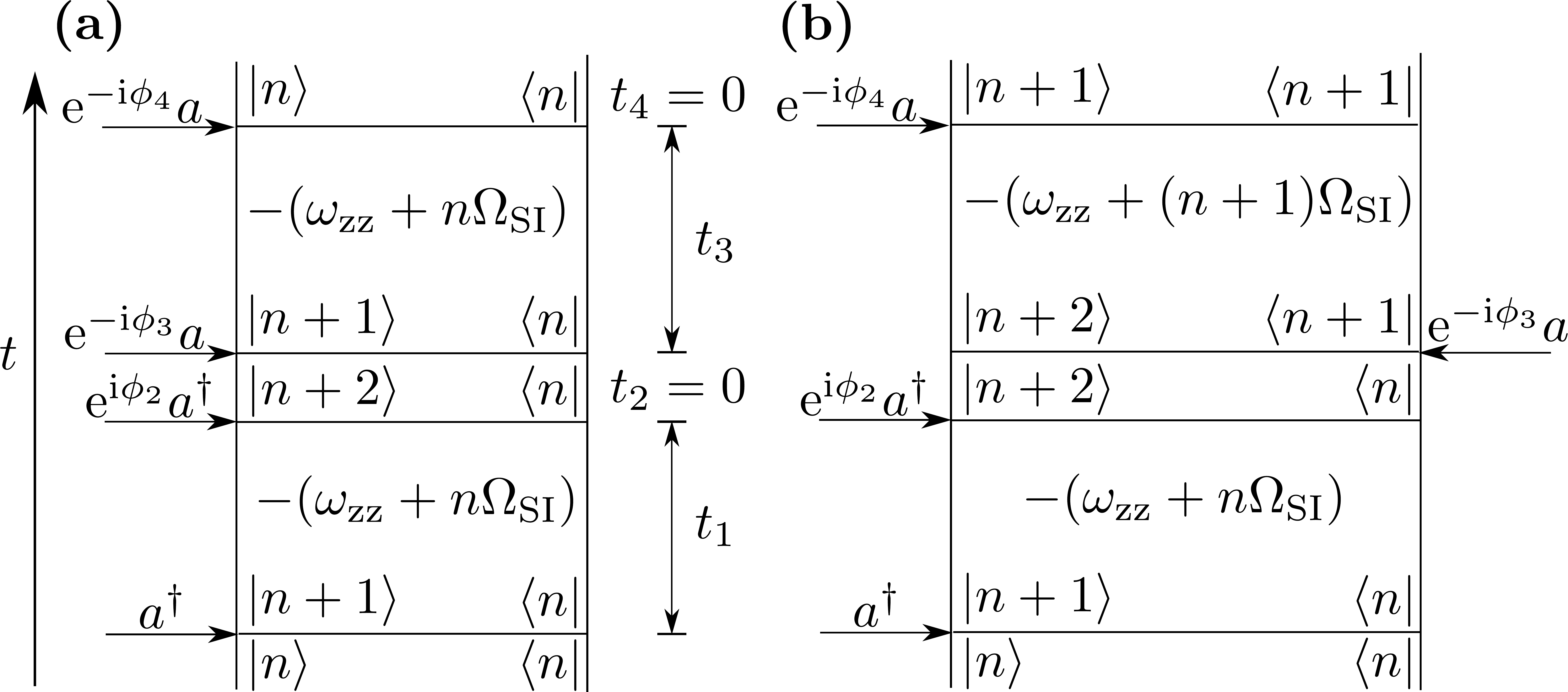}
\caption{Parts {\bf (a)} and {\bf (b)} of the figure show two example pathways carrying the
phase signature $\phi_2 -\phi_3 -\phi_4$. Starting from a population all pathways have to end in a population
in order to be observable. In paths {\bf (a)} the coherences oscillate with the same frequency
during the evolution period $t_3$ as during $t_1$ thus giving rise to diagonal peaks in the spectrum. 
In paths {\bf (b)} the oscillation frequency during $t_3$ is shifted vertically by $-\Omega_{\rm SI}$ with respect to $t_1$
leading to off-diagonal peaks below the main diagonal.}
\label{fig_pathways_fourth_order}
\end{figure}

In order to determine the self-interaction strength, we consider a sequence of four small displacements on the
zz-mode. A measurement of the zz-mode population completes the experimental cycle. We choose pathways
carrying the phase signature $\phi_2-\phi_3-\phi_4$; two example coherence transfer pathways are illustrated in Fig.~\ref{fig_pathways_fourth_order}. We
are interested in the dynamics during $t_1$ and $t_3$. Thus, our signal is given by Eqs. \eqref{eq:signal}
- \eqref{eq:rho} with $M=a_{\rm zz}^\dagger a_{\rm zz}$, $t_2=t_4=0$, $\alpha_k = |\alpha_k| e^{i\phi_k}$ for $k=1,2,3,4$ and $\phi_1=0$.

The practicality of our scheme is demonstrated by the simulation of the measurement of $\Omega_{\rm SI}$ for a realistic experimental
setting using $^{40}{\rm Ca}^+$ ions.
The motional states of the ions can be initialized close to their ground states by Doppler and sideband cooling~\cite{cooling_refs}
and the displacements of the modes can be implemented by state-dependent optical dipole forces~\cite{spin_dependent_forces}. 
Our parameters, summarized in Table~\ref{tab_fourth_order_parameters},
are sufficiently far from the structural transition so that a perturbative expansion remains valid and effective cooling
of the zigzag mode is still possible. 
We have not taken into account the effect of micromotion~\cite{normal_modes_micromotion}
which would lead to minor corrections of the entries of Table~\ref{tab_fourth_order_parameters} without affecting the general concepts
presented here.
In Table~\ref{tab_fourth_order_parameters} we also give the effective
dephasing rates for our choice of trap frequencies. For these parameters we expect dephasing
due to cross-Kerr couplings to be the dominant source of noise, so we neglect heating in our simulations.
The main contributions to the dephasing originate from the zigzag mode in $y$-direction and from
the Egyptian mode~\cite{sup_mat_a}, which we include in the simulations. We make $N_{\phi_k}=4$ phase cycles
for each phase and take all $|\alpha_k|=0.25$. We choose the initial state as a product of thermal states
for the modes with mean phonon numbers of $\bar{n}_{\rm zz}=1$ for the zigzag and $\bar{n}=4$ for the
other two modes. The motional Hilbert spaces are truncated including nine
energy levels for the zigzag and 15 for the other two modes which includes 99\% and 97\% of the respective
populations.

\begin{figure}
\includegraphics[width=\columnwidth]{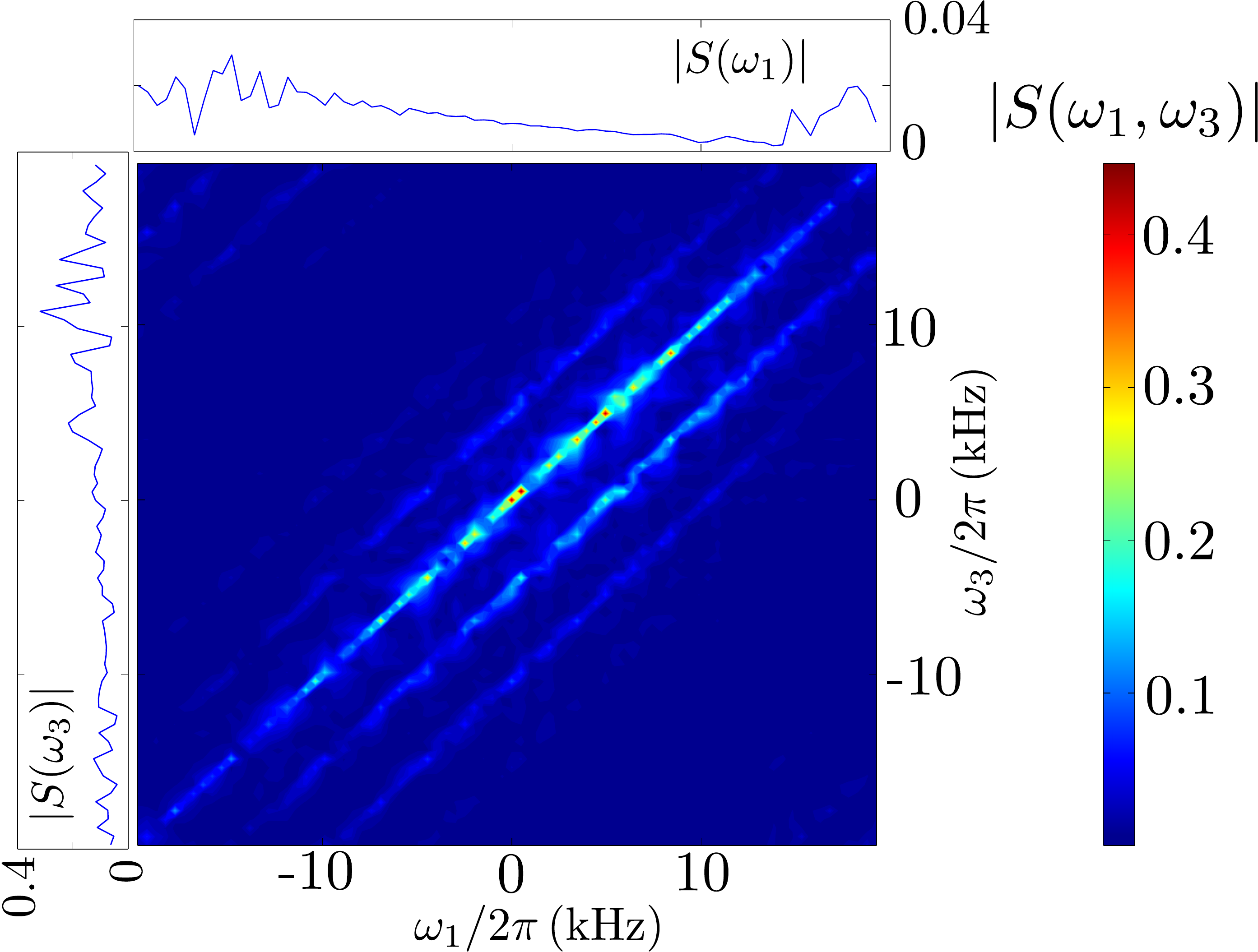}
\caption{The central plot shows the 2D spectrum $|S(\omega_1,\omega_3)|=|\mathcal{F}(s(t_1,t_3))|$ obtained by a four-pulse sequence with the
simulation parameters given in Table~\ref{tab_fourth_order_parameters}, including up to fourth-order terms in the
Hamiltonian, in the neighbourhood of the linear-to-zigzag transition. The diagonal peaks are due to paths of type {\bf (a)} in
Fig.~\ref{fig_pathways_fourth_order}. They are blurred because of static dephasing caused by thermal populations of the 
spectator modes leading to the diagonal line. The dominant off-diagonal (path {\bf (b)} in Fig.\ref{fig_pathways_fourth_order}) 
is shifted by $-\Omega_{\rm SI}$ along the $\omega_3$-axis and can thus be used to infer the self-interaction of the zigzag mode.
The small plots along $\omega_{1/3}$ show the spectra obtained by integrating along the other frequency direction.  This is the result that
would be obtained by a 1D experiment with only one free evolution period $|S(\omega_{1/3})|=|S(\omega_{1/3},t_{3/1}=0)|$. }
\label{fig_fourth_order_spectrum}
\end{figure}

The resulting 2D spectrum presented in Fig.~\ref{fig_fourth_order_spectrum} shows two dominant lines: one along the principal diagonal,
and one shifted below it. The principal diagonal is due to coherence transfer pathways where the coherences oscillate at the same 
frequency during $t_1$ and $t_3$. Example pathways are given in part {\bf (a)} of Fig.~\ref{fig_pathways_fourth_order}.
The off-diagonal line is due to paths where the oscillation frequency during $t_3$ is shifted by an amount $-\Omega_{\rm SI}$ 
with respect to the first free evolution period $t_1$, exemplified in part {\bf (b)} of Fig.~\ref{fig_pathways_fourth_order}.  
Therefore, this line shift gives direct access to the self-interaction strength $\Omega_{\rm SI}$.
In sharp contrast, a 1D-spectroscopy experiment with only one free evolution period would yield the 
information obtained by projecting the spectrum along one of the 
two frequency axes, so that $\Omega_{\rm SI}$ could not be obtained (cf. Fig.~\ref{fig_fourth_order_spectrum}).
Note that the coherence transfer pathways in Fig.~\ref{fig_pathways_fourth_order} would give rise to
a series of separated peaks; dephasing due to thermal occupation of the other modes blurs the maxima in the diagonal direction
giving rise to
the observed lines. All modes, except for the center-of-mass modes, contribute to this dephasing (though some of them quite weakly). 
Hence, by ground-state cooling of the modes contributing to dephasing one would obtain sharp and well-separated
resonances in the spectrum. This, however, is experimentally very demanding for large ion crystals.
Finally, we remark that phase fluctuations during the pulse sequence do not pose a problem for our protocol on the considered time scale.
For the use of optical dipole forces we estimate the loss in contrast due to laser phase fluctuations to be as little as 1\% for the signal
of the considered coherence transfer pathways~\cite{sup_mat_e}.

{\it Resonant energy exchange between normal modes investigated by 2D spectroscopy.} As a further example we consider a
parameter regime where the fourth-order terms are negligible and the dominant nonlinear effect in the dynamics is
coherent energy exchange between two modes due to a resonance in the third-order Hamiltonian $H^{(3)}$.
For a trap anisotropy $(\omega_z/\omega_x)^2 = 20/63$ we obtain a resonant coupling between the stretch
mode $c_2 = c_{\rm str}$ and the zigzag mode, of the form \cite{james_third_order_main}:
\beq
    H^{(3)}_{\rm res} = \hbar\Omega_{\rm T} [ a_{\rm zz}^2 c_{\rm str}^{\dagger} + (a_{\rm zz}^{\dagger})^2 c_{\rm str} ].
    \label{eff_third_order_ham}
\eeq
Here we have used a rotating-wave approximation in the frame rotating with the normal mode frequencies.
For the subspaces with the lowest phonon numbers, the eigenvectors and eigenvalues of $H^{(3)}_{\rm res}$
can be found analytically~\cite{sup_mat_g}; eigenvalues for higher occupation numbers may be
found numerically. We emphasize that a Hamiltonian up to third order is an approximation valid only for
low numbers of excitations, and fourth-order terms are necessary to guarantee a lower-bounded energy spectrum.

For an axial frequency $\omega_z/2\pi =2\,$MHz we obtain a coupling $\Omega_{\rm T}/2\pi=5.9\,$kHz. 
The nonlinear dynamics induced by $H^{(3)}_{\rm res}$ can be probed in a 2D experiment
with the same pulse sequence as described before, i.e. $|\alpha_k|=0.25$, $N_{\phi_k}=4$ and $t_{1/3}^{\rm max}= 2\,$ms,
reducing the time increment to $\Delta t_{1/3}= 10.6\,\mu$s. 
For our simulation parameters dephasing due to other modes is negligible and the dominant source of
decoherence is expected to be heating of the motional modes. Accordingly, we 
model the modes as damped harmonic oscillators coupled to thermal reservoirs at room temperature and assume
heating rates $\dot{\bar{n}}_{\rm zz/str} = 0.2/0.1\,{\rm quanta}\cdot{\rm ms}^{-1}$, a conservative estimate for
macroscopic traps~\cite{cooling_refs}. Furthermore, we take the initial state to be a product of thermal states
with residual phonon occupation numbers $\bar{n}_{\rm zz/str} = 0.7/0.2$. 
The Hilbert spaces are truncated at six and nine excitations for the stretch and zigzag modes, respectively, thus leaving out 
a fraction of $10^{-4}$ of the populations.

\begin{figure}[hbtp]
\includegraphics[width=0.85\columnwidth]{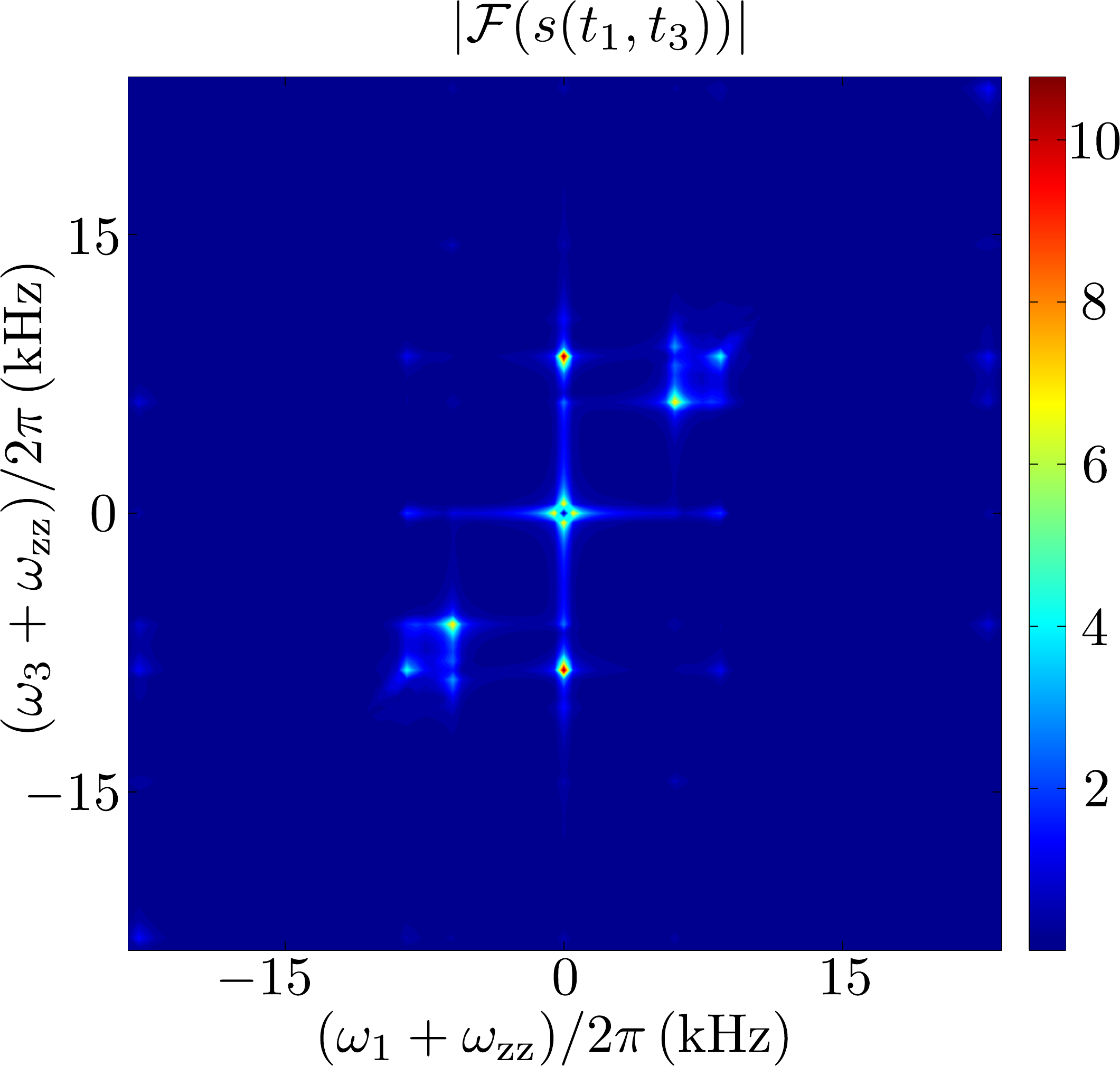}
\caption{2D spectrum due to the resonant third-order terms $H^{(3)}_{\rm res}$,~\eq{eff_third_order_ham}, in the Coulomb potential. Simulation parameters
are given in the main text. A strong peak at $\omega_1= \omega_3 = -\omega_{\rm zz}$
was removed from the spectrum for clarity. Eigenvalues of $H^{(3)}_{\rm res}$ for low phonon numbers are identified 
and the effect of homogeneous broadening is clearly visible as broadening of the peaks in vertical and horizontal
directions.}
\label{fig_third_order_spectrum}
\end{figure}

The resulting spectrum shown in Fig.~\ref{fig_third_order_spectrum} shows two bright peaks above and below the central peak, 
which correspond to pathways starting in the ground state.
Their vertical coordinates are shifted by $\pm \sqrt{2}\Omega_{\rm T}$, the eigenvalues
of $H^{(3)}_{\rm res}$ for the lowest levels showing coherent energy transfer between the two modes.
All peak coordinates are shifted with respect to $-\omega_{\rm zz}$ by an eigenvalue of $H^{(3)}_{\rm res}$
or a linear combination thereof, from which further eigenvalues can be inferred~\cite{sup_mat_g}. Off-diagonal
peaks, moreover, are an evidence of coherence transfer \cite{ernst_buch}. The figure clearly shows homogeneous broadening
of the peaks along the frequency axes due to the coupling to the thermal reservoirs. 
This illustrates how 2D spectroscopy allows for a distinction between homogeneous and inhomogeneous
broadening, since the latter leads to broadening of the peaks along the diagonal as in Fig.~\ref{fig_fourth_order_spectrum}.

In summary, we have shown how to extend 2D spectroscopy for the investigation of nonlinear dynamics of crystals of trapped
ions. The method offers significant advantages: it does not produce any signal for purely harmonic evolution
and it allows for the separation of signals which would appear superposed in a linear spectrum. It also facilitates
the characterization of noise in the system: while effective static disorder gives rise to diagonal lines, dephasing
and heating occuring during each experimental run manifest in broadening in the horizontal and vertical directions.
Furthermore, the protocol does not require ground-state cooling, a feature which is particularly appealing for
the study of large ion crystals. Note that it is well-known how to achieve significant reductions in the number 
of measurements required to obtain 2D spectra by employing techniques from the field of matrix completion~\cite{matrix_completion}.
The 2D spectroscopy methods presented here form a versatile new diagnostic toolbox
that may be applied well beyond the two case studies discussed here to cover all many-body models that may be
realized in ion traps including spin models, structural dynamics of large ion crystals, and models in which spin and vibrational degrees of freedom
are coupled.

{\em Acknowledgements.} The authors acknowledge discussions with U. Poschinger at early stages of
the project and useful comments on the manuscript from M. Bruderer. This work was supported
by the EU Integrating Project SIQS, the EU STREPs EQUAM and PAPETS, the Alexander von Humboldt Foundation and
the ERC Synergy Grant BioQ.

\newpage

\section{Supplemental Material to ``Two-dimensional spectroscopy for the study of ion Coulomb crystals''}

\maketitle

\tableofcontents

\appendix

\section{The motional Hamiltonian in the harmonic approximation}

We start by considering $N$ singly charged atomic ions of mass $m$ confined in a linear Paul trap. We assume the trapping potential to be harmonic such that we can write
\beq
V_{\rm t} = \sum_{i=1}^N \sum_{\mu=x,y,z} \frac{1}{2} m \omega_{\mu}^2 r_{i\mu}^2
\label{def_trapping_potential}
\eeq
where $\omega_{\mu}$ is the (pseudopotential) trapping frequency in spatial direction $\mu$ and $r_{i\mu}$ is the spatial coordinate $\mu$ of ion $i$.
The Coulomb interaction between the ions is given by:
\beq
V_{\rm C}=\frac{1}{2}\sum_{i,j\neq i} \frac{e^2}{4\pi \epsilon_0}\frac{1}{|{\bf r}_i-{\bf r}_j|}
\eeq
with $\epsilon_0$ the vacuum permittivity and $e$ the elementary charge. The full potential is then:
\beq
V=V_{\rm t}+V_{\rm C}.
\label{def_full_potential}
\eeq
Adding the kinetic energy of the ions we arrive at the full Hamiltonian for the motional degrees of freedom:
\beq
H= \sum_{i,\mu} \left( \frac{p_{i\mu}^2}{2 m} + \frac{1}{2} m \omega_{\mu}^2 r_{i\mu}^2 \right)
+ \frac{1}{2}\sum_{i \neq j} \frac{e^2}{4\pi \epsilon_0} \frac{1}{|{\bf r}_i - {\bf r}_j|}.
\label{full_ham_app}
\eeq
Assuming the trap axis along the spatial $z$-direction and $\omega_x,\:\omega_y \gg \omega_z$, the ions
arrange on a string along $z$ and the radial equilibrium positions are given by $x_i^0=y_i^0=0$ while the axial equilibrium positions are determined by
\beq
\left. \frac{\partial V}{\partial z_i} \right|_{z_i^0} = 0.
\label{z_eq_pos}
\eeq
Introducing the charateristic length scale
\beq
l_z^3=\frac{e^2}{4\pi \epsilon_0 m \omega_z^2}
\label{length_scale}
\eeq
the axial equilibrium positions $z_i^0$ may be written as
\beq
z_i^0 = l_z u_i^0
\label{scaled_eq_pos}
\eeq
 where the $u_i^0$  are usually termed the dimensionless equilibrium positions of the
ion string~\cite{james_normal_modes_sm}.~\eq{z_eq_pos} can then be written in terms of the $u_i^0$ which has the advantage
that the result is independent
of the specific ion mass and trapping frequency. The values of the $z_i^0$ for a specific setup are readily calculated with the help of $l_z$.
 A collection of the values of $u_i^0$ for up to ten ions may
be found in~\cite{james_normal_modes_sm}. If the ions are sufficiently cold they perform only small excursions  $\delta
r_{j \mu} (t)$
around their equilibrium positions such that their spatial coordinates can be expressed as
\beq
 r_{j \mu} (t) = r_{j \mu}^0 + \delta r_{j \mu} (t).
\eeq
Expanding the full potential in~\eq{def_full_potential} to second order in these small displacements from equilibrium
one obtains
\beq
V  \simeq V^{(2)} = \frac{1}{2} \left. \sum_{i,j,\mu,\beta} \frac{\partial^2 V}{\partial r_{i \mu} \partial r_{j \beta} }\right|_{{\bf r}_{i,j}^0} \delta r_{i \mu} \delta r_{j \beta}
\label{harmonic_approx_potential}
\eeq
where the constant energy shift due to the zeroth order contribution has been omitted.
For a
linear string of ions there are no couplings between the motion in different spatial directions in the second order approximation, so that the
potential is given by
\beq
V^{(2)} = \frac{m \omega_z^2}{2} \sum_{i,j} \sum_{\mu}  V^{\mu}_{ij} \delta r_{i\mu} \delta r_{j\mu}.
\label{second_order_potential}
\eeq
Before giving the expression for the $V^{\mu}_{ij}$ let us define the trap anisotropies
\beq
\alpha_{x/y}= \left( \frac{\omega_z}{\omega_{x/y}} \right)^2.
\eeq
Note that small values of $\alpha_{x/y}$ imply that the confinement in the radial directions is much tighter than in the axial direction.
With these definitions at hand we can write the Hessian matrices of the potential at the ions' equilibrium positions as~\cite{james_third_order_sm}
\beq
V_{ij}^z =  \begin{cases} 1 + 2\sum_{p \neq j} \frac{1}{|u_j^0-u_p^0|^3} & {\rm if} \hspace{1ex} i=j  \\
 \frac{-2}{|u_j^0-u_i^0|^3} & {\rm if}\hspace{1ex} i \neq j \end{cases}
\label{v_z}
\eeq
and
\beq
V_{ij}^{x/y} = \left(\frac{1}{\alpha_{x/y}} + \frac{1}{2} \right) \delta_{ij} - \frac{1}{2} V_{ij}^z.
\label{v_xy}
\eeq
Here, the $u_i^0$ are the dimensionless equilibrium positions defined in~\eq{scaled_eq_pos} and $\delta_{ij}$ is the Kronecker delta.

\begin{figure}
\includegraphics[width=\columnwidth]{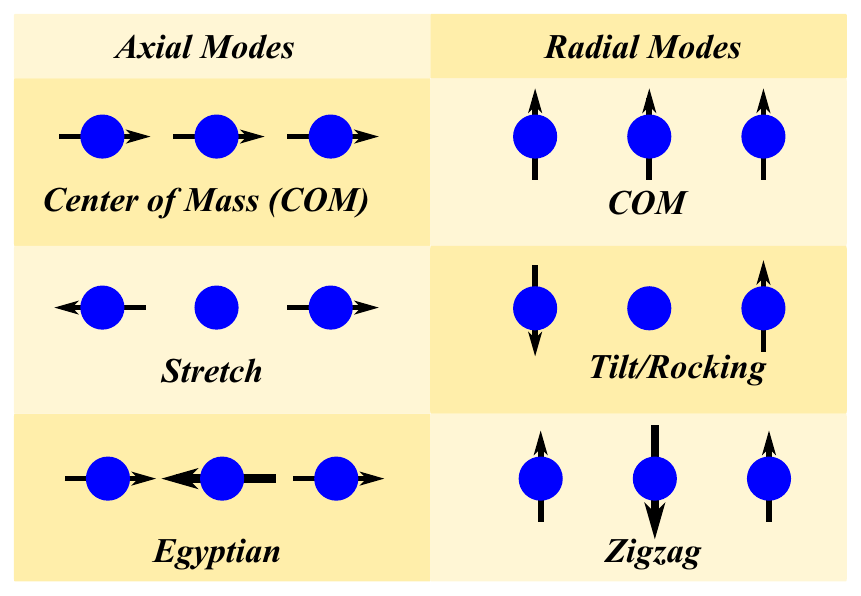}
\caption{Sketch illustrating the different vibrational eigenmodes of a linear three-ion crystal and the
standard naming convention for these modes. The inter-ion spacing is of the order of $l_z$.}
\label{fig_mode_structure}
\end{figure}

In the harmonic approximation the Hamiltonian~\eq{full_ham_app} may be written as
\beq
H = \sum_{i,\mu} \frac{p_{i\mu}^2}{2 m} + \frac{m \omega_z^2}{2} \sum_{i,j} \sum_{\mu}  V^{\mu}_{ij} \delta r_{i\mu} \delta r_{j\mu}.
\label{full_ham_app_harmonic_app}
\eeq
For each direction $\mu$, the matrix $V^{\mu}_{ij}$ is real and symmetric. As is apparent from Eqs.~\eqref{v_z}
and~\eqref{v_xy} all Hessians are diagonalized by the same orthogonal matrix $M$. The system is then described by $N$
uncoupled normal modes in every spatial direction~\cite{goldstein}.
Physically, the entries $M_{jn}$ of the matrix can be understood as the normalized amplitude of normal mode $n$ at ion $j$~\cite{goldstein,james_normal_modes_sm}.
Quantizing the motion according to
\beq
\begin{split}
r_{j\mu} & = \sum_n M_{jn} \sqrt{\frac{\hbar}{2 m \omega_{\mu,n}}} (a_{\mu,n} + a_{\mu,n}^{\dagger}), \\
p_{j\mu} & = \sum_n \ii M_{jn} \sqrt{\frac{\hbar m \omega_{\mu,n}}{2}} (a_{\mu,n}^{\dagger} - a_{\mu,n})
\end{split}
\label{def_quantization}
\eeq
the Hamiltonian~\eq{full_ham_app_harmonic_app} can be cast into the form
\beq
H_0= \sum_{\mu,n} \hbar \omega_{\mu,n} a_{\mu,n}^{\dagger} a_{\mu,n}
\label{def_H0}
\eeq
which corresponds to Eq.~(4) of the main text. Note that we have omitted the ground state energy.
The axial normal mode frequencies are given by
\beq
\omega_{z,n} = \sqrt{\lambda_{n}^{z}} \omega_z
\eeq
where $\omega_{z,n}$ is the frequency of mode $n$ and the $\lambda_n^z$ ($n=1,\dots,N$) are the eigenvalues of $V_{ij}^z$,
ordered such that they increase with increasing $n$.
Similarly, the radial normal mode frequencies read
\beq
\omega_{x/y,n} = \sqrt{\gamma_{n}^{x/y}} \omega_z
\eeq
where the $\gamma_{n}^{x/y}$ are the eigenvalues of $V_{ij}^{x/y}$ which can be written as
\beq
\gamma_{n}^{x/y} = \frac{1}{\alpha_{x/y}}+\frac{1}{2} - \frac{\lambda_n^z}{2}.
\label{radial_eigenvalues}
\eeq
Hence, the eigenvalues in the radial directions decrease with increasing $n$. Note, however, that the eigenvalues in the axial and radial directions with equal index $n$ correspond
to the same eigenvector and thus the corresponding modes have the same structure. 
It can be shown~\cite{james_normal_modes_sm} that the smallest eigenvalue of $V_{ij}^z$ is
$\lambda_1^z=1$
and corresponds to the center-of-mass mode. Thus, the center-of-mass mode is the energetically lowest lying mode in the
axial direction while it is the energetically highest lying mode
in the radial directions. Conversely, the energetically highest lying mode in the axial direction, corresponding to the
eigenvalue $\lambda_N^z$, is the mode where
neighboring ions perform out-of-phase oscillations. In the radial directions this mode is called the zigzag mode and is the energetically lowest lying mode.
In Fig.~\ref{fig_mode_structure} we schematically show all motional modes for a linear three-ion crystal.

\begin{figure}
\includegraphics[width=0.85\columnwidth]{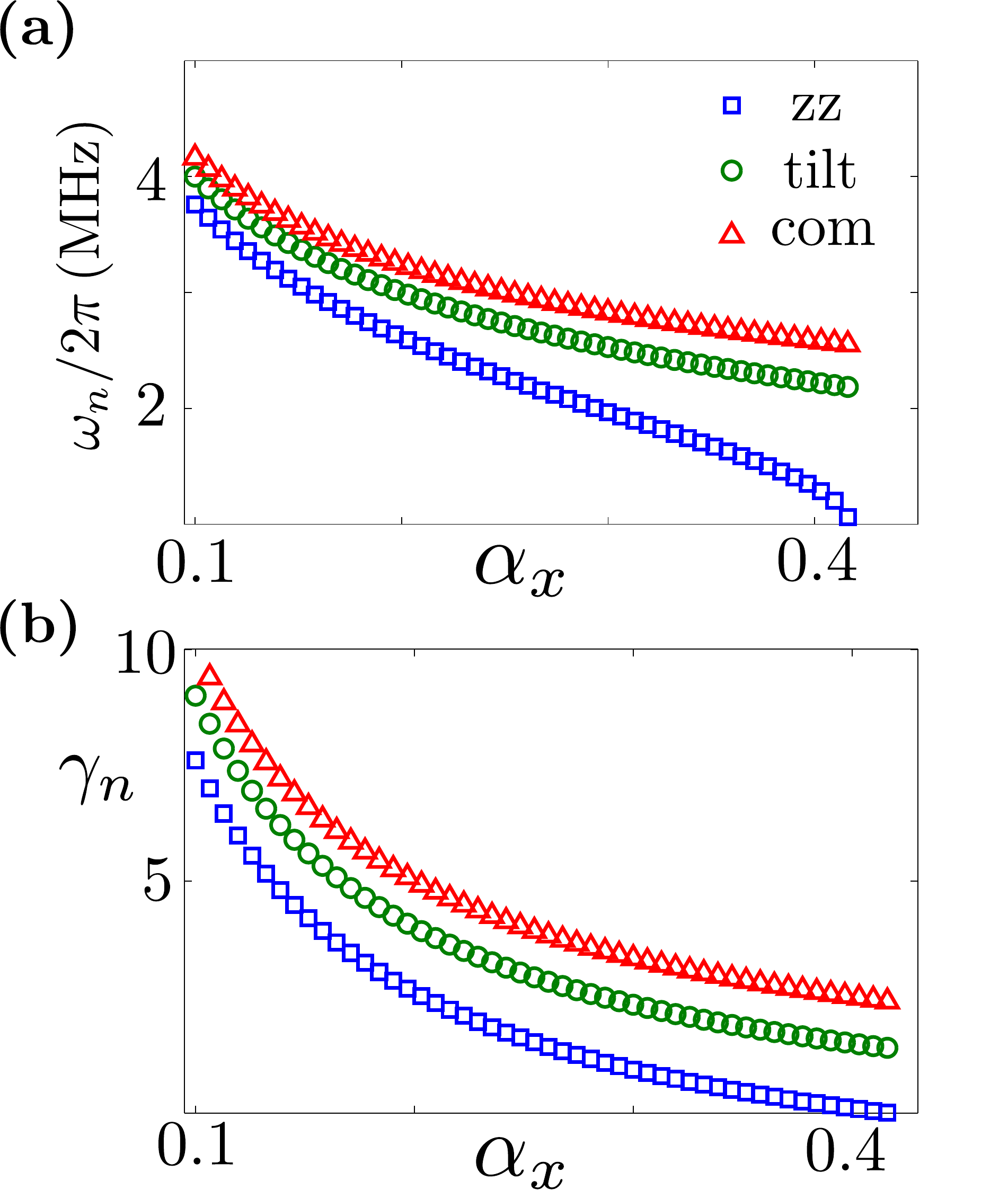}
\caption{ {\bf (a)}: Normal mode frequencies for a crystal of three $^{40}$Ca$^{+}$ ions as a function of the trap
anisotropy $\alpha_x=(\omega_z/\omega_x)^2$ for
an axial trapping frequency $\omega_z/2\pi=2\,$MHz. {\bf (b)}: Eigenvalues of the Hessian in $x$-direction as a function
of the trap anisotropy $\alpha_x$ for the
same parameters as in {\bf (a)}.}
\label{fig_mode_scaling}
\end{figure}

Approaching the linear-to-zigzag structural transition the zigzag mode is particularly relevant as it is the soft
mode in this transition for the case of infinitely many ions \cite{linear-zigzag_sm}
and numerical simulations show this behavior also emerges for chains of finite size.
The instability of the linear chain below a critical value of the transverse confinement can be seen
from~\eq{radial_eigenvalues}. If the radial confinement is lowered in one direction, e.g. the $x$-direction, while the
axial confinement is held constant, the value of $\alpha_x$ increases.
Accordingly, the eigenvalues $\gamma_n^x$ become smaller and for a certain anisotropy $\alpha_x$ we have
$\gamma_N^x=0$. Lowering the potential further would yield $\gamma_N^x<0$, an indicator that the linear configuration is
not stable anymore. Indeed, in this parameter regime the ions break away from
the linear configuration and arrange in a planar zigzag structure. This behavior is well-known and critical
anisotropies have been estimated and experimentally
tested for different numbers of ions \cite{linear-zigzag_experiment}. In Fig.~\ref{fig_mode_scaling} we show the scaling
of the eigenvalues and mode frequencies for the transverse modes in $x$-direction as a function
of the trap anisotropy for a crystal of three $^{40}$Ca$^{+}$ ions and an axial trapping frequency $\omega_z/2\pi=2\,$MHz.

\section{Third-order terms in the Coulomb potential}

In~\eq{def_trapping_potential} we have assumed that the trapping is harmonic which is a very good approximation in a
Paul trap, unless nonlinearities are deliberately added to the trapping potential, e.g. by means of optical fields
\cite{optical_potential}.
 The next order corrections to the potential in~\eq{harmonic_approx_potential} stem from the
Coulomb interaction and are obtained by evaluating
\beq
V^{(3)} = \frac{1}{3!} \sum_{i,j,k \atop \mu,\beta,\gamma} \left. \frac{\partial^3 V_{\rm C}}{\partial r_{k\gamma} \partial r_{i\mu} \partial r_{j\beta} } \right|_{{\bf r}_{i,j,k}^0}  \delta r_{i\mu} \delta r_{j\beta} \delta r_{k\gamma}.
\label{third_order_general}
\eeq

Inserting the dimensionless equilibrium positions from~\eq{scaled_eq_pos} into the above expression, after some algebra
we obtain~\cite{james_third_order_sm}
\beq
V^{(3)} = \frac{m \omega_z^2}{2 l_z} \sum_{i,j,k} C_{ijk} (3 \delta x_i \delta x_j + 3 \delta y_i \delta y_j - 2 \delta z_i \delta z_j)\delta z_k
\eeq
with the tensor $C_{ijk}^{(3)}$
\beq
C^{(3)}_{ijk}= \begin{dcases}
\sum_{p\neq k} \frac{{\rm sgn}(u_k^0 - u_p^0)}{|u_k^0-u_p^0|^4} & {\rm if}\: i=j=k \\
\frac{{\rm sgn}(u_k^0 - u_j^0)}{|u_k^0-u_j^0|^4} & {\rm if}\: i=j\neq k \\
\frac{{\rm sgn}(u_i^0 - u_k^0)}{|u_i^0-u_k^0|^4} & {\rm if}\: i\neq j=k \\
\frac{{\rm sgn}(u_j^0 - u_k^0)}{|u_j^0-u_k^0|^4} & {\rm if}\: i=k\neq j \\
0 & {\rm if} i\neq j\neq k. \\
\end{dcases}
\label{def_cijk3}
\eeq
Quantizing the coordinates according to~\eq{def_quantization} and using $a_{x,n}=a_n,\:a_{y,n}=b_{n}$ and $a_{z,n}=c_n$
we arrive at the third-order correction Hamiltonian~\cite{james_third_order_sm}
\beq
\begin{split}
H^{(3)}_{\rm full} =& \frac{z_0}{4 l_z}\hbar \omega_z \sum_{n,m,p} D_{nmp}^{(3)} \times \\
& \left[\frac{3 (a_n + a_n^{\dagger}) (a_m+a_m^{\dagger})}{\sqrt[4]{\gamma_n^x \gamma_m^x}} +\frac{3(b_n + b_n^{\dagger}) (b_m+b_m^{\dagger})}{\sqrt[4]{\gamma_n^y \gamma_m^y}}   \right.\\
&\left. - \frac{2 (c_n + c_n^{\dagger}) (c_m+c_m^{\dagger})}{\sqrt[4]{\lambda_n^z \lambda_m^z}}  \right] \frac{( c_p +c_p^{\dagger})}{\sqrt[4]{\lambda_p^z}}  \\
\end{split}
\label{full_quantized_third_order_ham}
\eeq
where $D_{nmp}^{(3)}$ is defined as
\beq
D_{nmp}^{(3)}=\sum_{i,j,k} C_{ijk}^{(3)} M_{in} M_{jm} M_{kp}
\eeq
and $z_0=\sqrt{\hbar/(2 m \omega_z)}$ is the spread of the longitudinal center-of-mass ground state wave packet. We note
that all $D_{nmp}^{(3)}$ involving the center-of-mass modes, i.e. all $D_{nmp}^{(3)}$ with at least one index equal to one, vanish. 
Physically, this is due to the fact that transferring excitations to or from the center-of-mass modes would change the momentum of the
crystal as a whole. This, however, cannot be due to the Coulomb interaction~\cite{james_third_order_sm}.

Transforming the quantized interaction Hamiltonian~\eq{full_quantized_third_order_ham} to an interaction picture with
respect to the time-independent Hamiltonian $H_0$,~\eq{def_H0}, it can
be inspected for resonances. This has been done in detail in~\cite{james_third_order_sm},
where a list of possible resonances, i.e. involved modes and according values of $\alpha_{x/y}$, is provided for
up to $N=10$ ions. Two types of resonances can occur: a radial phonon is created while one longitudinal and one radial
phonon are annihilated or two transverse phonons are created upon the annihilation of one longitudinal phonon.
Considering the case
of $N=3$ ions only the latter resonance is possible. For our simulations we assumed that the degeneracy between the radial modes is lifted such that a resonance involves only one radial mode and the
other mode is sufficiently far off-resonant. We chose to consider the $x$-direction for the transverse modes. For $\alpha_x=20/63$ one obtains a resonant coupling between the $x$-zigzag mode and the
stretch mode. In this case the resonant part of the Hamiltonian in~\eq{full_quantized_third_order_ham} reads
\beq
H_{\rm res}^{(3)}= 3 \frac{z_0}{4 l_z} \hbar \omega_z \frac{D_{332}^{(3)}}{\sqrt[4]{\gamma_3^x \gamma_3^x \lambda_2^z}} \left[ a_3^2 c_2^{\dagger} + (a_3^{\dagger})^2 c_2 \right].
\eeq
Setting $\Omega_{\rm T}= 3 z_0 \omega_z D_{332}^{(3)}/(4 l_z \sqrt[4]{\gamma_3^x \gamma_3^x \lambda_2^z})$ as well as $a_3=a_{\rm zz},\:c_2=c_{\rm str}$
we recover Eq.~(9) of the main text.

\section{Fourth-order terms in the Coulomb potential}

The fourth-order corrections to the potential in~\eq{harmonic_approx_potential} due to the Coulomb
potential are obtained by evaluating the expression
\beq
V^{(4)} = \frac{1}{4!} \sum_{i,j,k,l \atop \mu,\beta,\gamma,\delta} \left. \frac{\partial^4 V_{\rm C}}{\partial r_{l\delta} \partial r_{k\gamma} \partial r_{i\mu} \partial r_{j\beta} } \right|_{{\bf r}_{i,j,k,l}^0}
\delta r_{i\mu} \delta r_{j\beta} \delta r_{k\gamma} \delta r_{l\delta}.
\label{fourth_order_general}
\eeq
After a rather lengthy but straightforward calculation we arrive at
\beq
\begin{split}
V^{(4)} = &\frac{1}{4!} \frac{3 e^2}{4 \pi \epsilon_0} \frac{1}{l_z^5} \sum_{i,j,k,l} C_{ijkl}^{(4)} \left[ 3 \delta x_i \delta x_j \delta x_k \delta x_l  \right. \\
& + 3 \delta y_i \delta y_j \delta y_k \delta y_l    + 8 \delta z_i \delta z_j \delta z_k \delta z_l + 6 \delta x_i \delta x_j \delta y_k \delta y_l \\ 
& \left. - 24 \delta x_i \delta x_j \delta z_k \delta z_l - 24  \delta y_i \delta y_j \delta z_k \delta z_l \right]
\end{split}
\label{}
\eeq
where
\beq
\begin{split}
C_{ijkl}^{(4)} = \delta_{ij} \delta_{jk} \delta_{kl} \sum_{p\neq l} \frac{1}{|u_l^0 -u_p^0|^5} +  \delta_{ij} \delta_{jk} (1 -\delta_{kl}) \frac{-1}{|u_k^0 -u_l^0|^5} \\
+ \delta_{ij} (1-\delta_{jk}) \delta_{lj} \frac{-1}{|u_l^0 -u_k^0|^5} + (1-\delta_{ij}) \delta_{jk} \delta_{kl} \frac{-1}{|u_l^0 -u_i^0|^5} \\
+ (1-\delta_{ij}) \delta_{ik} \delta_{kl} \frac{-1}{|u_j^0 -u_l^0|^5} + \delta_{ij} (1-\delta_{jk}) \delta_{kl} \frac{1}{|u_j^0 -u_l^0|^5} \\
 + (1-\delta_{ij}) \delta_{jk} \delta_{il} \frac{1}{|u_k^0 -u_l^0|^5} + (1-\delta_{ij}) \delta_{ik} \delta_{jl} \frac{1}{|u_l^0 -u_k^0|^5}
\end{split}
\label{def_cijkl_full}
\eeq
or
\beq
C_{ijkl}^{(4)} = \begin{dcases}
\sum_{p\neq l} \frac{1}{|u_l^0 -u_p^0|^5} & {\rm if}\:\: i=j=k=l \\
 \frac{-1}{|u_k^0 -u_l^0|^5} & {\rm if}\:\: i=j=k\neq l \\
 \frac{1}{|u_j^0 -u_l^0|^5} & {\rm if}\:\: i=j\neq k=l \\
0 & {\rm else.}
\end{dcases}
\label{def_cijkl_short}
\eeq
The indices of $C_{ijkl}^{(4)}$ can be interchanged at will. Thus, the above definition covers all entries of the
tensor. Using~\eq{def_quantization} we obtain the quantized form of the fourth-order expansion of the Coulomb potential.
The calculation is straightforward and yields
\beq
\begin{split}
& H_{\rm full}^{(4)} = \left( \frac{z_0}{4 l_z} \right)^2 \hbar \omega_z \sum_{n,m,p,q} D_{nmpq}^{(4)} \times \\
& \left[ \frac{3}{\sqrt[4]{\gamma_n^x \gamma_m^x \gamma_p^x \gamma_q^x}} (a_n+a_n^{\dagger}) (a_m+a_m^{\dagger}) (a_p+a_p^{\dagger}) (a_q+a_q^{\dagger}) \right. \\
& +\frac{3}{\sqrt[4]{\gamma_n^y \gamma_m^y \gamma_p^y \gamma_q^y}} (b_n+b_n^{\dagger}) (b_m+b_m^{\dagger}) (b_p+b_p^{\dagger}) (b_q+b_q^{\dagger})  \\
& +\frac{8}{\sqrt[4]{\lambda_n^z \lambda_m^z \lambda_p^z \lambda_q^z}} (c_n+c_n^{\dagger}) (c_m+c_m^{\dagger}) (c_p+c_p^{\dagger}) (c_q+c_q^{\dagger})  \\
& +\frac{6}{\sqrt[4]{\gamma_n^x \gamma_m^x \gamma_p^y \gamma_q^y}} (a_n+a_n^{\dagger}) (a_m+a_m^{\dagger}) (b_p+b_p^{\dagger}) (b_q+b_q^{\dagger})  \\
& -\frac{24}{\sqrt[4]{\gamma_n^x \gamma_m^x \lambda_p^z \lambda_q^z}} (a_n+a_n^{\dagger}) (a_m+a_m^{\dagger}) (c_p+c_p^{\dagger}) (c_q+c_q^{\dagger})  \\
& \left. -\frac{24}{\sqrt[4]{\gamma_n^y \gamma_m^y \lambda_p^z \lambda_q^z}} (b_n+b_n^{\dagger}) (b_m+b_m^{\dagger}) (c_p+c_p^{\dagger}) (c_q+c_q^{\dagger})  \right]
\end{split}
\label{full_quantized_fourth_order_ham}
\eeq
where we have introduced
\beq
D_{nmpq}^{(4)} = \sum_{i,j,k,l} C_{ijkl}^{(4)} M_{in} M_{jm} M_{kp} M_{lq}.
\eeq

Again, there are no couplings to the center-of-mass modes. This can be shown following the proof for the third order in
Ref.~\cite{james_third_order_sm}. We start by realizing that
\beq
\sum_l C_{ijkl}^{(4)} = 0.
\eeq
This can be seen in the following way: if $i\neq j \neq k$ all terms in the above sum are zero and the result is
trivial; if $i =j \neq k$ we have $\sum_l C_{iikl}^{(4)} = C_{iiki}^{(4)} + C_{iikk}^{(4)}=0$
as $C_{iiki}^{(4)} = -C_{iikk}^{(4)}$ from the definition in~\eq{def_cijkl_short}; in case $i=j=k$ we have
$\sum_l C_{iiil}^{(4)} = C_{iiii}^{(4)}+ \sum_{l\neq i} C_{iiil}^{(4)}$
which is again found to be zero by using~\eq{def_cijkl_short}. Using that $M_{l1}=1/\sqrt{N},\:l=1,\dots,N$ we obtain
\beq
\begin{split}
D_{nmp1}^{(4)} &= \sum_{i,j,k,l} C_{ijkl}^{(4)} M_{in} M_{jm} M_{kp} \frac{1}{\sqrt{N}} \\
&=\frac{1}{\sqrt{N}} \sum_{i,j,k}  M_{in} M_{jm} M_{kp}  \sum_l C_{ijkl}^{(4)} = 0.
\end{split}
\eeq
As the indices of $C_{ijkl}^{(4)}$ can be interchanged freely this is true for every element of $D_{nmpq}^{(4)}$ with at least one index equal to 1.

Let us now identify the regimes in which the fourth-order terms in the
Hamiltonian,~\eq{full_quantized_fourth_order_ham}, have appreciable contributions to the motional dynamics.
Under normal trapping conditions ($\omega_x,\omega_y\gg \omega_z$) the corrections to the harmonic
Hamiltonian,~\eq{def_H0}, are very small due to the prefactor $[z_0/(4 l_z)]^2 \approx 10^{-6}$. For instance,
in~\cite{innsbruck_cross_kerr_sm} Kerr-type interactions due to the fourth-order terms of the Coulomb potential were found
to have strengths $\sim 1-10\,$Hz. The situation is different approaching the linear-to-zigzag
transition. If the trapping potential in $x$-direction is relaxed reaching the close vicinity of the structural
transition, $\gamma_N^x$ approaches zero while all other
eigenvalues have values well above zero (cf. Fig.~\ref{fig_mode_scaling} for the case of three ions). Hence, due to the appearance of $\gamma_N^x$ in the denominator, the terms involving $\gamma_N^x$ acquire
an appreciable value. Retaining only terms involving modes in the $x$-direction in~\eq{full_quantized_fourth_order_ham} yields the Hamiltonian in Eq.~(6)
of the main text.

\begin{table*}
\caption{\bf Shifts in the normal mode frequencies due to fourth-order effects of the Coulomb interaction for a crystal of
$N=3$ ions with $\omega_{\rm z}/2\pi=2\,{\rm MHz}$ and
$\omega_{\rm x}/2\pi=3.1012\,{\rm MHz},\:\omega_{\rm y}/2\pi=5\,{\rm MHz}$}
\renewcommand{\arraystretch}{1.4}
\begin{tabular}{p{2cm}p{2cm}p{2cm}p{2cm}p{2cm}p{2cm}p{2cm}}
\hline
\hline
Frequency shift & $\Delta \omega_{\rm x,2}/2\pi$ & $\Delta \omega_{\rm zz}/2\pi$ & $\Delta \omega_{\rm y,2}/2\pi$ &
$\Delta \omega_{\rm y,3}/2\pi$ & $\Delta \omega_{\rm z,2}/2\pi$ & $\Delta \omega_{\rm z,3}/2\pi$ \\
\hline
Third order & -0.5008$\,$kHz &   -10.0850 $\,$kHz &    0$\,$kHz  &  0$\,$kHz & 0.5275 $\,$kHz &   0.2821  $\,$kHz          \\
\hline
Fourth order &    0.4791$\,$kHz &   25.2874$\,$kHz &   0.0826$\,$kHz &    0.2894$\,$kHz  & -0.4371$\,$kHz &
-0.9430$\,$kHz       \\
\hline
Effective &   -0.0217$\,$kHz &   15.2025$\,$kHz &     0.0826$\,$kHz &    0.2894$\,$kHz &  0.0905$\,$kHz &
-0.6609$\,$kHz   \\
\hline
\hline
\end{tabular}
\label{tab_freq_shifts}
\end{table*}

\begin{table*}
\caption{\bf Dephasing rates due to fourth-order effects of the Coulomb interaction  for a crystal of $N=3$ ions with
$\omega_{\rm z}/2\pi=2\,{\rm MHz}$ and
$\omega_{\rm x}/2\pi=3.1012\,{\rm MHz},\:\omega_{\rm y}/2\pi=5\,{\rm MHz}$}
\renewcommand{\arraystretch}{1.4}
\begin{tabular}{p{2cm}p{2cm}p{2cm}p{2cm}p{2cm}p{2cm}p{2cm}}
\hline
\hline
 & $\Omega_{\rm d,2}^x/2\pi$ & $(\Omega_{\rm SI}/2)/2\pi$ & $\Omega_{\rm d,2}^y/2\pi$ & $\Omega_{\rm d,3}^y/2\pi$ &
$\Omega_{\rm d,2}^z/2\pi$ & $\Omega_{\rm d,3}^z/2\pi$ \\
\hline
Third order &  -1.0487$\,$kHz &  -10.3467$\,$kHz &  0$\,$kHz &    0$\,$kHz &   1.0551$\,$kHz &     0.5171$\,$kHz     \\
\hline
 Fourth order &   0.9582$\,$kHz &    12.9082$\,$kHz &   0.1652$\,$kHz &     0.5787$\,$kHz &  -0.8741$\,$kHz &
-1.8860$\,$kHz     \\
\hline
Effective &  -0.0905$\,$kHz &     2.5615$\,$kHz &   0.1652$\,$kHz &     0.5787$\,$kHz &  0.1810$\,$kHz &
-1.3690$\,$kHz      \\
\hline
\hline
\end{tabular}
\label{tab_dephasing_rates}
\end{table*}

This Hamiltonian has to be considered in more detail in order to identify the contributions relevant for the dynamics.
To facilitate the analysis we divide the Hamiltonian in Eq.~(6) of the main text into three parts $H^{(4)}= H_{\rm xx}
+H_{\rm xy}+H_{\rm xz}$ where the indices denote the spatial direction in which the first and last two pairs of
operators act. We restrict our analysis to the case of $N=3$ ions; the generalization to larger $N$ is straightforward.
The first term reads
\beq
\begin{split}
H_{\rm xx} = &3 \left( \frac{z_0}{4 l_z} \right)^2 \hbar \omega_z \sum_{n,m,p,q}  \frac{D_{nmpq}^{(4)}}{\sqrt[4]{\gamma_n^x \gamma_m^x \gamma_p^x \gamma_q^x}} \times \\
& (a_n+a_n^{\dagger}) (a_m+a_m^{\dagger}) (a_p+a_p^{\dagger}) (a_q+a_q^{\dagger}).
\end{split}
\eeq
Moving to a frame rotating with the phonon frequencies we can find the resonant terms which will contribute appreciably
to the dynamics. The Hamiltonian contains many Kerr-type terms $a_n^{\dagger} a_n a_m^{\dagger} a_m$ (and permutations
thereof) coupling two modes $n$ and $m$. These terms do not acquire a time dependence in the
rotating frame. In the special case $m=n$ these terms can be written as $(a_n^{\dagger})^2 a_n^2 + a_n^{\dagger} a_n$.
Thus, the non-rotating fourth-order terms lead to Kerr-type couplings between different modes
and also to a self-interaction of the modes together with a shift of the mode frequencies. In order to make sure that
only these non-rotating terms are the dominant contribution to the dynamics
one needs to check if the time-dependent terms can be neglected in a RWA. To this end we assume realistic
experimental parameters as summarized in Table~I of the main text and perform a quantitative
anlysis:
For each combination $nmpq$ there are 16 operator terms $O_{nmpq,i},\:i=1,\dots,16$, whose coefficients and
frequencies we denote by $c_{nmpq,i}$ and $\omega_{nmpq,i}$, respectively. Then, we check
if $c_{nmpq,i}/\omega_{nmpq,i} \ll 1$ for all energy non-conserving terms where $\omega_{nmpq,i}\neq 0$. This analysis shows that the energy non-conserving terms can be neglected in a RWA.
Finally, ordering the resonant terms yields
\beq
\begin{split}
H_{\rm xx} =& \frac{\hbar \Omega_{\rm SI}}{2} (a_{\rm zz}^{\dagger})^2 a_{\rm zz}^2 + \hbar \Delta \omega_{{\rm zz}} a_{\rm zz}^{\dagger} a_{\rm zz} \\
&+ \hbar \Omega_{\rm d,2}^x  a_{\rm zz}^{\dagger} a_{\rm zz} a^{\dagger}_2 a_2 + \hbar \Delta \omega_{x,2} a^{\dagger}_2 a_2
\end{split}
\label{h_xx}
\eeq
where
\beq
\begin{split}
\Omega_{\rm SI} &= 12\times 3 \left( \frac{z_0}{4 l_z} \right)^2  \frac{D_{3333}^{(4)}}{\gamma_{\rm zz}^x} \omega_z, \\
\Omega_{\rm d,2}^x &= 24 \times 3 \left( \frac{z_0}{4 l_z} \right)^2  \frac{D_{2233}^{(4)}}{\sqrt{\gamma_2^x \gamma_{\rm zz}^x}} \omega_z, \\
\Delta \omega_{\rm zz} &= \Omega_{\rm SI} + \frac{1}{2}\Omega_{\rm d,2}^x, \\
\Delta \omega_{x,2} &= \frac{1}{2}\Omega_{\rm d,2}^x \\
\end{split}
\label{correction_definitions2}
\eeq
and we use the index zz (instead of 3) to denote the $x$ zigzag mode. Note the we omitted the self-interaction of the $x$ tilt mode as it does not involve $\gamma_{\rm zz}^x$.
Numerical values for the frequency shifts and mode couplings for the parameters used in our simulations can be found in
Tables~\ref{tab_freq_shifts} and~\ref{tab_dephasing_rates}.

Following the same procedure we obtain
\beq
H_{\rm xy} = a_{\rm zz}^{\dagger} a_{\rm zz} \left(\hbar \Delta \omega_{\rm zz}' + \sum_{n=2,3} \hbar \Omega_{\rm d,n}^{y}   b_n^{\dagger} b_n \right) +  \sum_{n=2,3} \hbar \Delta \omega_{y,n} b_n^{\dagger} b_n
\label{h_xy}
\eeq
where
\beq
\begin{split}
\Omega_{\rm d, 2}^y &= 4 \times 6 \left( \frac{z_0}{4 l_z} \right)^2  \frac{D_{3322}^{(4)}}{\sqrt{\gamma_{\rm zz}^x \gamma_2^y}} \omega_z, \\
\Omega_{\rm d, 3}^y &= 4 \times 6 \left( \frac{z_0}{4 l_z} \right)^2  \frac{D_{3333}^{(4)}}{\sqrt{\gamma_{\rm zz}^x \gamma_3^y}} \omega_z, \\
\Delta \omega_{\rm zz}' &= \frac{1}{2}(\Omega_{\rm d, 2}^y + \Omega_{\rm d, 3}^y), \\
\Delta \omega_{y,n} &= \frac{\Omega_{\rm d, n}^y}{2}.\\
\end{split}
\label{correction_definitions}
\eeq
Again numerical values can be found in Tables~\ref{tab_freq_shifts} and~\ref{tab_dephasing_rates}. Note that the frequency shift of the $x$ zigzag
mode given in Table~\ref{tab_freq_shifts} is the sum of the shifts in Eqs. \eqref{h_xx}, \eqref{h_xy} and \eqref{h_xz}.
The resonant contributions of the third part are given by
\beq
H_{\rm xz} = a_{\rm zz}^{\dagger} a_{\rm zz}  \left( \hbar \Delta \omega_{\rm zz}'' + \sum_{n=2,3} \hbar \Omega_{\rm d,n}^{z}   c_n^{\dagger} c_n \right) + \sum_{n=2,3} \hbar \Delta \omega_{z,n} c_n^{\dagger} c_n
\label{h_xz}
\eeq
where the mode couplings $\Omega_{\rm d,n}^{z}$ and the frequency shifts $\Delta \omega_{z,n}$ are defined analogous as in~\eq{correction_definitions}. 
Considering the numerical values for the mode couplings and frequency shifts in Tables~\ref{tab_freq_shifts} and~\ref{tab_dephasing_rates}
we realize that, except for the $x$ zigzag mode, the frequency shifts are much smaller than the motional frequencies which are of the order of MHz.
Therefore, we neglect the frequency shifts for all modes except the zigzag mode. The sum of the above Hamiltonians 
$H^{(4)}= H_{\rm xx} +H_{\rm xy}+H_{\rm xz}$ then yields the structure of the Hamiltonian $H_{\rm eff}^{(4)}$ in Eqs.~(7) 
and~(8) of the main text. Let us emphasize that the self-interaction and frequency shift of the
zigzag mode increase fastest when approaching the structural transition, scaling as $1/\gamma_{\rm zz}^x$, while
the mode couplings only scale as $1/\sqrt{\gamma_{\rm zz}^x}$ (cf. Eqs.~\eqref{correction_definitions2} and~\eqref{correction_definitions}).

An analysis including only the fourth-order terms is incomplete, since the off-resonant third-order Hamiltonian terms
induce energy shifts of the same order of magnitude as the fourth-order terms, leading to corrections for the
self-interactions,
frequency shifts and dephasing rates~\cite{innsbruck_cross_kerr_sm, NRJ_2009_sm}. For
our purposes we only need to consider the part of the Hamiltonian $H^{(3)}_{\rm full}$,~\eq{full_quantized_third_order_ham},
involving $x$ modes, namely
\beq
\begin{split}
H^{(3)}_{\rm x}= &3 \frac{z_0}{4 l_z} \hbar \omega_z \sum_{n,m,p} \frac{D_{nmp}^{(3)}}{\sqrt[4]{\gamma_n^x \gamma_m^x \lambda_p^z}} \times \\
& (a_n + a_n^{\dagger}) (a_m+a_m^{\dagger}) ( c_p +c_p^{\dagger}).
\end{split}
\label{third_order_ham}
\eeq
Hence, the off-resonant energy shifts induced by third-order terms are obtained from
\beq
\Delta E = \sum_{ \{ n'\} \neq \{n \} } \frac{|\bra{\{ n'\}} H^{(3)}_{\rm x} \ket{\{n \}}|^2}{E_n-E_{n'}}
\label{third_order_corrections_app}
\eeq
where $\ket{\{n \}}$ is a motional Fock state with energy $E_n$ and $\ket{\{ n'\}}$ is any other motional Fock state with
energy $E_{n'}$. Realizing that $D^{(3)}_{233},\:D^{(3)}_{323}$ and $D^{(3)}_{332}$ are the only non-zero elements of
$D^{(3)}_{nmp}$ involving the $x$ zigzag mode means that only states that differ in the quantum numbers $n_{\rm
x,2},n_{\rm x,3},n_{\rm z,2},n_{\rm z,3}=n_{\rm t},n_{\rm zz},n_{\rm str},n_{\rm eg}$ are coupled and contribute to the energy
shifts. More precisely, one finds that the state $\ket{\{ n\}}=\ket{n_{\rm t},n_{\rm zz},n_{\rm str},n_{\rm eg}}$ is coupled 
to the states $\ket{\{ n'\}}=\ket{n_{\rm t}\pm 1,n_{\rm zz} \pm 1,n_{\rm str},n_{\rm eg} \pm
1}$, $\ket{\{ n'\}}=\ket{n_{\rm t},n_{\rm zz}\pm 2,n_{\rm str}\pm 1,n_{\rm eg}}$ and
$\ket{\{ n'\}}=\ket{n_{\rm t},n_{\rm zz},n_{\rm str}\pm 1,n_{\rm eg}}$ by $H^{(3)}_{\rm x}$.
 Note that in the states where various ``$\pm$'' appear
all possible combinations of plus and minus signs are relevant. The numerical values for the coupling rates and frequency shifts obtained from~\eq{third_order_corrections_app}
are summarized in Tables~\ref{tab_freq_shifts} and~\ref{tab_dephasing_rates}.

Finally, the effective coupling rates and frequency shifts are obtained by summing the third- and fourth-order
contributions and can be found in the last rows of Tables~\ref{tab_freq_shifts} and~\ref{tab_dephasing_rates}. The
relative frequency shifts due to higher-order corrections of the Coulomb potential are
$10^{-4}$ or smaller for all modes except the $x$ zigzag mode and are therefore neglected for these modes. From Eqs.~\eqref{h_xy} and~\eqref{h_xz} we see that a thermal occupation
of the $y$ and $z$ modes, with the exception of the center-of-mass modes, causes shifts of the $x$ zigzag mode frequency and leads to effective dephasing. 
Therefore, we refer to the mode couplings $\Omega_{\rm d,n}^{\mu}$ also as dephasing rates. The numbers  in
Table~\ref{tab_dephasing_rates} show that the coupling of the $x$ zigzag mode to other modes is strongest for the $y$ zigzag and the Egyptian mode. 
Accordingly, a thermal population of these modes yields the strongest contribution to the effective
dephasing of the $x$ zigzag mode. The other dephasing contributions are considerably smaller. Combining these considerations with Eqs.~\eqref{h_xx},~\eqref{h_xy} and~\eqref{h_xz} 
we arrive at the effective Hamiltonian used for the simulations presented in the main text
\begin{equation}
\begin{split}
H_{\rm eff}^{(4)} =& \frac{\hbar \Omega_{\rm SI}}{2} (a^{\dagger}_{\rm zz})^2 a_{\rm zz}^2 + \hbar \Delta \omega_{\rm zz} a_{\rm zz}^{\dagger}
a_{\rm zz} \\
&+ a_{\rm zz}^{\dagger} a_{\rm zz} \left( \hbar \Omega_{\rm d,3}^{y}  b_3^{\dagger} b_3 + \hbar \Omega_{\rm d,3}^{z} c_3^{\dagger}
c_3\right).
\end{split}
\end{equation}

\section{Coherence transfer pathways and phase cycling}

\begin{figure}
\includegraphics[width=0.9\columnwidth]{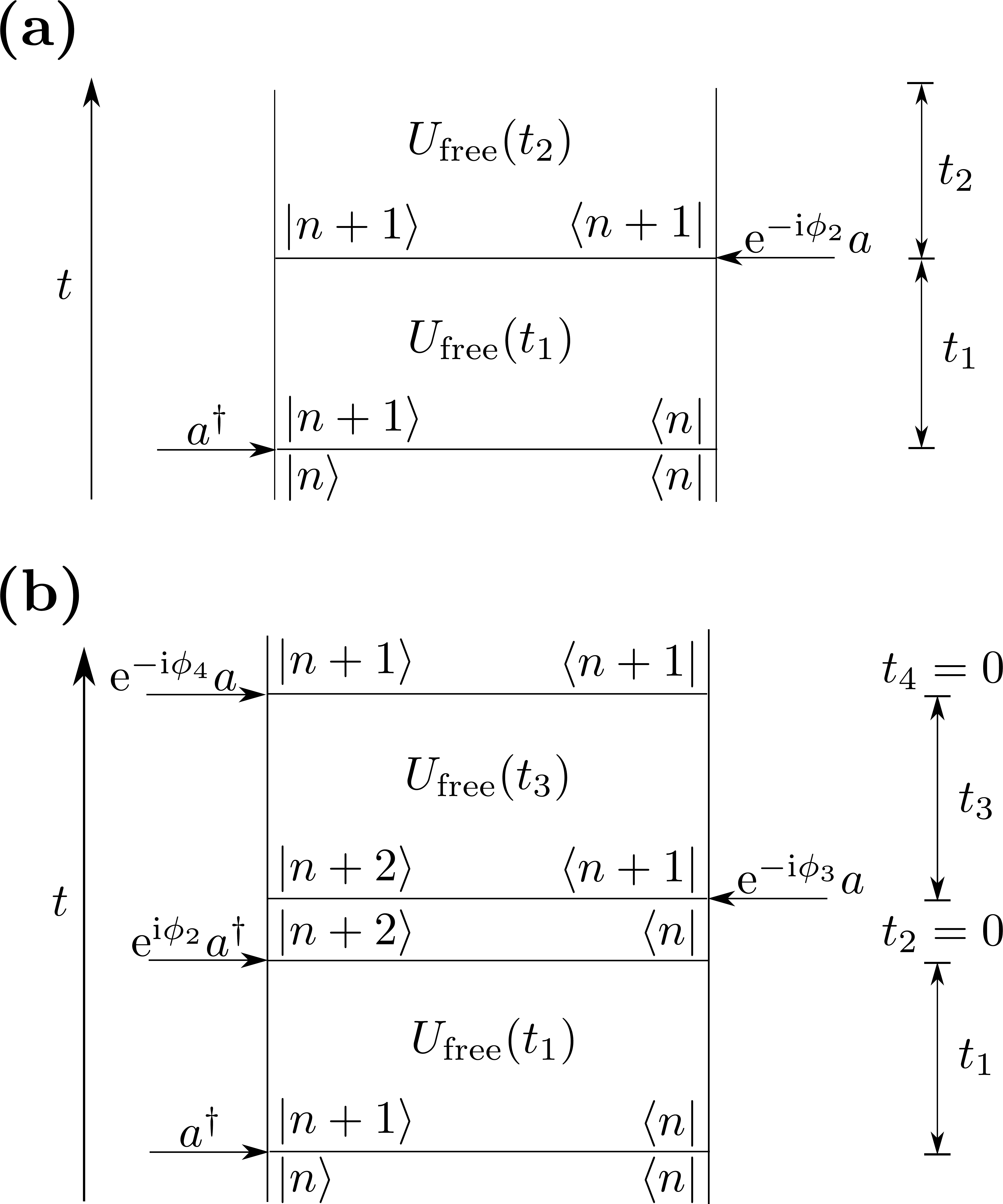}
\caption{ {\bf (a):} A coherence transfer pathway for the simplest pulse sequence,~\eq{time_ev_simple_pathway}, yielding
a two-dimensional spectrum in our setup.
The population $\ketbra{n}$ is transferred to the observable population $\ketbra{n+1}$. The phase $-\phi_2$ is imprinted
on the final population.
 {\bf (b):} A coherence transfer pathway for an experiment consisting of four displacements and two free evolutions is
shown.
The phase signature $\phi_2-\phi_3-\phi_4$ is imprinted onto the final population.  }
\label{fig_cpt_illustration}
\end{figure}

In this section we briefly summarize the ideas of {\it coherence transfer pathways} and {\it phase cycling}~\cite{ernst}
which provide a means of postselecting only a certain set of contributions to
the signal measured in a 2D experiment. We follow the treatment in~\cite{ernst}.
Let us consider the simplest 2D experiment with trapped ions following the scheme introduced in
the main text consisting of two displacement pulses followed by free evolutions. The time-evolution operator reads
\beq
U_{0} (t_1,t_2) = U_{\rm free} (t_2) D(\alpha_2) U_{\rm free} (t_1) D(\alpha_1).
\label{time_ev_simple_pathway}
\eeq
The concept of coherence transfer pathways in this context is the following: each of the displacements is written as a
Taylor series in powers of $\alpha_k$ and we pick only one operator term acting on the density matrix from each
side for each displacement. Every possible combination of these terms forms a coherence transfer pathway. However, only
pathways that end up in a population contribute to the measured signal.
A simple example of a pathway for an experiment described by the sequence in~\eq{time_ev_simple_pathway}
is illustrated in Fig.~\ref{fig_cpt_illustration} {\bf (a)}.

In the illustrated sequence, the first displacement sets a phase reference such that all subsequent displacements can be
applied with well-defined phase, in this case $\alpha_2 = |\alpha_2| \ee^{\ii \phi_2}$. The phase of the second
displacement is ``imprinted'' on the final state; for example, the contribution to the signal from the pathway shown in
Fig.~\ref{fig_cpt_illustration} {\bf (a)} is proportional to $\ee^{-\ii \phi_2}$ and therefore we say it carries a phase
signature $-\phi_2$. There are other pathways with different phases that also end in a diagonal matrix element.
The observable we measure in the end is the mode population $\langle a^{\dagger}a \rangle = \langle \hat{n} \rangle$.
If $c_n$ is the probability of state $\ket{n}$ in the final state of the experiment, the signal is given by
\beq
\begin{split}
s(t_1,t_2,\phi_2) &= \langle \hat{n} \rangle = \sum_n c_n n  = \sum_n n \sum_p c_{n,p} \ee^{\ii p \phi_2} \\
& =\sum_{p} \ee^{\ii p \phi_2} \sum_n c_{n,p} n= \sum_{p} s_p (t_1,t_2) \ee^{\ii p \phi_2}.
\end{split}
\label{derivation_signal}
\eeq
For future reference we summarize the above equation to
\beq
s(t_1,t_2,\phi_2)=\sum_{p} s_p (t_1,t_2) \ee^{\ii p \phi_2}.
\label{signal_decomposition}
\eeq
Note that the result is a real number. We now want to obtain the contribution to the signal 
due to pathways like the one illustrated in Fig.~\ref{fig_cpt_illustration} {\bf (a)}. This means
we need to obtain the contribution to the signal carrying the phase signature $q \phi_2$
where $q =-1$.
In order to achieve this, one performs $N_{\phi_2}$ experiments varying the phase $\phi_2$ systematically as
\beq
\phi_{2,k} = k \frac{2\pi}{N_{\phi_2}},\hspace{2ex}k=0,\dots,N_{\phi_2}-1.
\eeq
The signal obtained from each of these experiments is made up of a superposition as in~\eq{signal_decomposition}. Thus, we can obtain the contribution that stems from the pathways
with the phase signature $-\phi_2$ by a discrete Fourier analysis of the signal obtained in the $N_{\phi_2}$ experiments
\beq
\begin{split}
s(t_1,t_2,q) &= \frac{1}{N_{\phi_2}} \sum_{k=0}^{N_{\phi_2}-1}  s(t_1,t_2,\phi_{2,k})  \ee^{-\ii q \phi_{2,k}} \\
&= \frac{1}{N_{\phi_2}} \sum_{k=0}^{N_{\phi_2}-1} \left[ \sum_{p} s_p(t_1,t_2) \ee^{\ii p \phi_{2,k}} \right] \ee^{-\ii q \phi_{2,k}}.
\end{split}
\eeq
This procedure is called phase cycling.

However, in this way one does not only obtain the contribution with the phase factor $q \phi_2$ but also all
contributions with $(q + r N_{\phi_2})\phi_2$ where
 $r \in \mathbb{Z}$. Thus, after phase cycling the signal is made up of all selected contributions
\beq
s(t_1,t_2,q) = s_{q}  (t_1,t_2) + \sum_r s_{q + rN_{\phi_2}}  (t_1,t_2).
\label{full_selected_signal}
\eeq
The unwanted but selected contributions in the second part of~\eq{full_selected_signal} are due to
terms of higher orders in the $\alpha_i$ in the expansion of the displacements. Hence, one must choose $|\alpha_i|$
sufficiently small so that only the first few terms of the Taylor expansion are non-negligible. Then, for large enough
$N_{\phi_2}$ only the desired pathway contributes to the signal. This defines what we mean  by a ``small'' displacement
for a given $N_{\phi_2}$. Note that phase cycling requires an increase in the number of experiments performed by a factor $N_{\phi_2}$.

This procedure can be generalized to more than one phase. In Fig.~\ref{fig_cpt_illustration} {\bf (b)} we illustrate a pathway for a sequence of four displacements
interleaved by two free evolutions as in the experiments proposed in the main text. The full time evolution operator then reads
\beq
U_{0} (t_1,t_3) = D(\alpha_4) U_{\rm free} (t_3) D(\alpha_3)D(\alpha_2) U_{\rm free} (t_1) D(\alpha_1)
\eeq
where we have also set $t_2=t_4=0$. Here we can write
\beq
\alpha_i = |\alpha_i| \ee^{\ii \phi_i}
\eeq
for $i \geq 2$. In the case illustrated in Fig.~\ref{fig_cpt_illustration} {\bf (b)} the phase $\phi_2-\phi_3-\phi_4$ is imprinted
on the final population $\ketbra{n+1}$. If we are interested in selecting
the pathways with this phase signature we can perform phase cycling for each of these phases performing $N_{\phi_i}$ experiments for each phase.
Defining $ {\bf q} = (q_2,q_3,q_4)=(1,-1,-1)$ and $\phi_{\bf k}=(\phi_{2,k_2},\phi_{3,k_3},\phi_{4,k_4})$ we obtain
the desired signal through
\beq
s(t_1,t_3, {\bf q}) = \frac{1}{N_{\rm tot}} \sum_{k_2}\sum_{k_3}\sum_{k_4} s(t_1,t_3,\phi_{\bf k})  \ee^{-\ii {\bf q} \cdot \phi_{\bf k}}
\eeq
where $N_{\rm tot}=N_{\phi_2} \cdot N_{\phi_3} \cdot N_{\phi_4}$. Note that this increases the number of experiments
one needs to perform by a factor $N_{\rm tot}$.

\section{Impact of phase fluctuations on phase cycling}

In the previous section it was shown how to select only a certain set of contributions to the signal of a 2D spectroscopy experiment by 
means of phase cycling. Phase cycling relies on the ability to apply a series of displacements on the considered motional mode with  
well-defined phases. So far, we have assumed that these phases can be controlled arbitrarily well. Yet, this is not
true in a real experiment. In fact, in any real experiment the phases of the applied displacement pulses will be subject to fluctuations.
In the following we shall analyze how these phase fluctuations affect the signal obtained in a 2D spectroscopy
experiment. We will consider that the displacements 
are implemented by a state-dependent optical dipole force on the ions induced by laser radiation~\cite{monroe_science_sm}. 

We start by considering the simplest case of an experiment involving only two displacements as described in the first part of the previous section.
In this case the signal is ideally given by (see~\eq{signal_decomposition})
\beq
s(t_1,t_2,\phi_2)=\sum_{p} s_p (t_1,t_2) \ee^{\ii p \phi_2}.
\label{ideal_signal_decomposition}
\eeq
In practice, however, the signal considered in the above equation is the mean obtained from a series of experiments.
In every experimental run the phase of the second pulse is subject to some small fluctuations about the desired value such that
the phase $\phi_2$ becomes a random variable. Hence,~\eq{ideal_signal_decomposition} becomes
\beq
s(t_1,t_2,\phi_2) =\sum_{p} \langle s_{p} (t_1,t_2) \ee^{\ii p \phi_{2}} \rangle_{\rm st}
\label{signal_stochastic_average_a}
\eeq
where $\langle \dots \rangle_{\rm st}$ denotes the stochastic average. We are interested in the impact of phase fluctuations on the signal with a certain phase signature 
$q\phi_2$. Therefore, we want to compute the corrections to the signal $s(t_1,t_2,\phi_2)$ caused by the fluctuating phase. 
We start by noting that the $s_p (t_1,t_2)$ are independent of the phase $\phi_2$ (see~\eq{derivation_signal}). 
Further, we assume that for the $m$th experimental run we can write $\phi_{2,m} = \phi_{2} + \Delta \phi_{2,m}$ 
where $|\Delta \phi_{2,m}| \ll 1$, i.e. that the phase fluctuations are small on the considered timescale. This seems justified in light of the results of~\cite{phase_fluctuations_fsk}.
There it was found that laser phase drifts of $2\pi$ occured on a timescale $\tau_{\rm d} \approx 10\,$s while the experiments we consider take place on a
timescale $\tau_{\rm exp} \approx 5\,$ms. We can then write ~\eq{signal_stochastic_average_a} as
\beq
s(t_1,t_2,\phi_2) =\sum_{p} \ee^{\ii p \phi_2} s_{p} (t_1,t_2) \langle \ee^{\ii p \Delta \phi_{2,m}} \rangle_{\rm st}.
\label{signal_stochastic_average_b}
\eeq
In order to proceed we assume that the laser phase drifts can be modeled as a Wiener process~\cite{stochastic_processes_gillespie}. 
A Wiener process $X(t)$ is a Gaussian process that obeys the stochastic differential equation
\beq
\frac{{\rm d} X(t)}{{\rm d} t} = \sqrt{c} \Gamma(t)
\eeq
where $c>0$ and $\Gamma(t)$ is Gaussian white noise. $X(t)$ is characterized by its initial value $X(t_0)$ and the diffusion constant $c$. Its first and second moments read
\beq
\langle X(t) \rangle = X(0),\hspace{2ex} {\rm Var}[X(t)] = ct
\eeq
where have set the initial time $t_0=0$. The covariance of the Wiener process for two times $s,t$ is given by~\cite{breuer_petruccione}
\beq
{\rm Cov}[X(s),X(t)] = c \cdot {\rm min}(s,t).
\label{cov_wiener_process}
\eeq
For the phase fluctuations we assume a Wiener process with zero mean and diffusion constant $c=(4\pi^2/10)\cdot {\rm s}^{-1}$. We determine $c$ 
by identifying the standard deviation of the process at $t=10\,$s with $2\pi$. $\Delta \phi_{2,m}$ is then given by the value of the stochastic 
process at the instance of time when the second laser pulse is applied. Using $|\Delta \phi_{2,m}| \ll 1$ we can simplify~\eq{signal_stochastic_average_b}
to
\beq
\begin{split}
s(t_1,t_2,\phi_2) &\approx \sum_{p} \ee^{\ii p \phi_2} s_p (t_1,t_2) \left(1 - \frac{1}{2} p^2 {\rm Var}[\Delta \phi_{2,m}] \right) \\
& = \sum_{p} \ee^{\ii p \phi_2} s_p (t_1,t_2) \left(1 - \frac{1}{2} p^2 c \tau_{\rm exp} \right).
\end{split}
\label{corrected_signal}
\eeq
Here we have already used that $\langle \Delta \phi_{2,m}\rangle = 0$. Using $\tau_{\rm exp} = 5\,$ms and $c=(4\pi^2/10)\cdot {\rm s}^{-1}$ as introduced 
above we obtain corrections of about 1\% for terms with $p=1$ and 4\% for $p=2$. 

We now turn to the case of a protocol including four displacements as proposed in the main text where $t_2=t_4=0$.
The above result is readily extended to this case. The signal for the protocol we propose in the main text is ideally given by
\beq
\begin{split}
s(t_1,t_3,\boldsymbol{\phi}) &=\sum_{\bf p} s_{\bf p} (t_1,t_3) \ee^{\ii {\bf p} \cdot \boldsymbol{\phi}} \\
&= \sum_{p_2,p_3,p_4} s_{p_2,p_3,p_4} (t_1,t_3) \ee^{\ii p_2 \phi_2} \ee^{\ii p_3 \phi_3} \ee^{\ii p_4 \phi_4}.
\end{split}
\label{ideal_signal_four_pulses}
\eeq
We again assume that each of the phases may be written as $\phi_{i,m}=\phi_i + \Delta \phi_{i,m},\:i=2,3,4$. Note, however, that for a specific $m$ the fluctuations
are not uncorrelated as they are samples of the same stochastic process at different instances of time. We can then write~\eq{ideal_signal_four_pulses} as
\beq
s(t_1,t_3,\boldsymbol{\phi}) =\sum_{\bf p} \ee^{\ii {\bf p} \boldsymbol{\phi}} s_{{\bf p}} (t_1,t_3) \langle \ee^{\ii p_2 \Delta \phi_{2,m}} \ee^{\ii p_3 \Delta \phi_{3,m}} \ee^{\ii p_4 \Delta \phi_{4,m}} \rangle_{\rm st}. 
\label{average_four_pulses_a}
\eeq
Again we have used that $s_{{\bf p}} (t_1,t_3)$ is independent of the phase fluctuations.
We simplify the above equation by expanding the exponentials to second order in the small phase fluctuations.
Using the covariance property,~\eq{cov_wiener_process}, of the Wiener process we can cast~\eq{average_four_pulses_a} into the form
\beq
\begin{split}
s(t_1,t_3,\boldsymbol{\phi}) &\approx \sum_{\bf p} \ee^{\ii {\bf p} \boldsymbol{\phi}} s_{\bf p} (t_1,t_3) (1 - \frac{1}{2} c (p_2+p_3+p_4)^2 t_1 -\frac{1}{2} c p_4^2 t_3) \\
&=\sum_{\bf p} \ee^{\ii {\bf p} \boldsymbol{\phi}} s_{\bf p} (t_1,t_3) (1 - \Delta s_{\bf p}(t_1,t_3))
\end{split}
\label{contrast_four_pulses}
\eeq
where we have used that the fluctuations have a zero mean and $t_2=0$. 
Based on~\eq{contrast_four_pulses}  we can now estimate the loss in the signal, and thus contrast, for different pathways. In order to provide an upper bound
for the loss in signal we set $t_1=t_3=2.5\,$ms for our estimates. In Tab.~\ref{tab_loss} we summarize the values for $\Delta s_{\bf p}(t_1,t_3)$ for pathways whose signals $\propto |\alpha|^4$ and 
$\propto |\alpha|^6$. For the pathway $(p_2,p_3,p_4)=(1,-1,-1)$ which we chose for our simulations the loss in signal is about 1\%. Thus, laser phase fluctuations do not
pose a substantial problem for the protocol. In fact, as can be seen in Table~\ref{tab_loss} the losses lie in the range of 1-5\% for all considered pathways. Signal contributions which
scale with higher powers of $|\alpha|$ are negligible in view of the smallness of $|\alpha|$.

\begin{table}
\caption{\bf Loss in signal due to laser phase fluctuations for different pathways of 2D experiment including four displacements}
\begin{tabular}{p{25pt}p{25pt}p{25pt}p{25pt}}
\hline
\hline
$ p_2$ & $ p_3$ & $ p_4$ & $\Delta s_{\bf p}(t_1,t_3)$ \\
\hline
1 & -1 & -1 & 1.0\% \\
\hline
1 & -2 & -1 & 2.5\% \\
\hline
1 & -1 & -2 & 4.0\% \\
\hline
2 & -2 & 1 & 1.0\% \\
\hline
-1 & -1 & -1 & 5.0\% \\
\hline
\hline
\end{tabular}
\label{tab_loss}
\end{table}

\section{Cancellation of the signal from harmonic systems}

In this section we will show that there is no signal for purely harmonic systems in an experiment using the four pulse
sequence we propose in the main text. For clarity, we start by considering only the addressed mode and the case of unitary free
evolution, and then show how to extend the result to systems with several modes and in contact with thermal baths. The
full time-evolution operator for the experiments we propose reads
\beq
U_{0} (t_1,t_3) = D(\alpha_4) U_{\rm free} (t_3) D(\alpha_3)D(\alpha_2) U_{\rm free} (t_1) D(\alpha_1).
\label{full_time_ev_supp_mat}
\eeq
We assume that the displacements can be effectively written as
\beq
D(\alpha_i) = \mathds{1} + |\alpha_i| (\ee^{\ii \phi_i} a^{\dagger} - \ee^{-\ii \phi_i} a )
\label{displacement_expansion}
\eeq
where higher powers in $\alpha_i$ are either cancelled by phase cycling or give a negligible contribution to the spectrum as $|\alpha_i| \ll 1$.

The signal at the end of each experimental cycle is given by
\beq
s(t_1,t_3) = \tr [\rho(t_1,t_3) n]
\label{signal_supp_mat}
\eeq
where $n$ is the number operator of the mode which is displaced in the experimental sequence. Using $\rho(t_1,t_3) = U_{0}(t_1,t_3) \rho_0 U_{0}^{\dagger}(t_1,t_3)$
we may write~\eq{signal_supp_mat} as
\beq
s(t_1,t_3) = \tr[\rho_0 U_{0}^{\dagger} (t_1,t_3) n U_{0}(t_1,t_3)]
\eeq
where $U_{0} (t_1,t_3)$ is defined in~\eq{full_time_ev_supp_mat} and $\rho_0$ is the initial state of the phonon modes. We now focus on the expression
$U_{0}^{\dagger} (t_1,t_3) n U_{0}(t_1,t_3)$ and define $U_1$ such that $U_0 (t_1,t_3) = D(\alpha_4) U_1 $. Using the expansion in~\eq{displacement_expansion}
and keeping only terms with the phase $\ee^{-\ii \phi_4}$ we obtain
\beq
\begin{split}
U_{0}^{\dagger} (t_1,t_3) n U_{0}(t_1,t_3) &= |\alpha_4| \ee^{-\ii \phi_4} U_1^{\dagger} [a,n] U_1 \\
& = |\alpha_4| \ee^{-\ii \phi_4} U_1^{\dagger} a U_1.
\end{split}
\eeq
Next we write $U_1$ as $U_1 = U_{\rm free}(t_3) U_2$. Note that for a harmonic time evolution we have $U_{\rm free}^{\dagger} (t_3) a U_{\rm free}(t_3)= f(a,a^{\dagger})$
where $f(a,a^{\dagger})$ is a function {\it linear} in $a$ and $a^{\dagger}$. Thus, we obtain
\beq
U_{0}^{\dagger} (t_1,t_3) n U_{0}(t_1,t_3) =|\alpha_4| \ee^{-\ii \phi_4}  U_2^{\dagger} f(a,a^{\dagger}) U_2.
\label{first_time_ev_signal}
\eeq
We now substitute $U_2 = D(\alpha_3) U_3$ and insert the expression into~\eq{first_time_ev_signal}. Again using the expansion
in~\eq{displacement_expansion} and keeping only terms with the phase $\ee^{-\ii \phi_3}$ we obtain
\beq
\begin{split}
U_{0}^{\dagger} (t_1,t_3) n U_{0}(t_1,t_3) &=|\alpha_3||\alpha_4| \ee^{-\ii (\phi_3+\phi_4)} U_3^{\dagger} [a,f(a,a^{\dagger})] U_3\\
& = |\alpha_3||\alpha_4| \ee^{-\ii (\phi_3+\phi_4)} U_3^{\dagger} c\mathds{1} U_3.
\end{split}
\eeq
As $f(a,a^{\dagger})$ is linear in the creation and destruction operators we have $[a,f(a,a^{\dagger})] =c \mathds{1}$ with some $c\in\mathds{C}$.
We then introduce $U_3=D(\alpha_2) U_4$. Inserting this expression in the above equation and following the same procedure as above
we finally obtain
\beq
\begin{split}
U_{0}^{\dagger} (t_1,t_3) n U_{0}(t_1,t_3) &=-|\alpha_2||\alpha_3||\alpha_4| \ee^{\ii (\phi_2-\phi_3-\phi_4)} U_4^{\dagger} [a^{\dagger},c \mathds{1}] U_4 \\
&= 0.
\end{split}
\eeq
Thus, we obtain no signal for purely harmonic time evolution.

The same calculation can be generalized to the case in which there are additional modes in the system. For this, one
uses that harmonic evolution maps generalized quadrature operators, i.e. linear combinations of creation and
annihilation operators for the different modes, into other quadrature operators, and that the commutator of two
generalized quadratures is proportional to the identity. This line of reasoning can also be extended to systems in
contact with thermal baths, or subject to other non-unitary dynamics leading to linear evolution. This is done
by including the environmental degrees of freedom within the system, so that the total evolution becomes unitary and the
above argument can be applied.

\section{Identification of peaks in the third-order spectrum}

In this section we will identify the peaks that appear in the spectrum shown in Fig.~3 of
the main text.
To this end let us briefly recall that the spectrum is due to dynamics induced by the third-order corrections of the
Coulomb potential.
In the particular case considered the dynamics is governed by the Hamiltonian in Eq.~(9) of the main text, namely
\beq
H^{(3)}_{\rm res} = \hbar \Omega_{\rm T} (a_{\rm zz}^2 c_{\rm str}^{\dagger} + (a_{\rm zz}^{\dagger})^2 c_{\rm str}).
\label{eff_third_order_ham_app}
\eeq
Note once again that this Hamiltonian is not bounded from below and therefore only valid in the regime of low phonon
numbers or, equivalently, small
oscillation amplitudes. For high excitation numbers the fourth-order terms must be taken into
account.

In the regime of low phonon numbers one can find a few of the eigenvectors and eigenvalues of the third-order
Hamiltonian in~\eq{eff_third_order_ham_app} analytically. We start by realizing that $H^{(3)}_{\rm res}$ commutes with the
operator $n_{\rm zz} + 2 n_{\rm str}$. Taking into account that $\omega_{\rm str}=2\omega_{\rm zz}$ we see that $H^{(3)}_{\rm res}$ 
only induces transitions between states which are degenerate with respect to the harmonic Hamiltonian
in~\eq{def_H0}. In Table~\ref{tab_eigenvalues} the eigenvalues of $H^{(3)}_{\rm res}$ can be found together with the Fock states 
of which the eigenstates are linear combinations.

\begin{table}[bt]
\caption{\bf First five eigenvalues and corresponding manifolds of $H^{(3)}_{\rm res}$}
\renewcommand{\arraystretch}{1.5}
\begin{tabular}{ccc}
 Manifold $\ket{n_{\rm str}, n_{\rm zz}}$ &$\phantom{\hspace{10pt}}$&  Eigenvalues \\
$\ket{1,0},\ket{0,2}$ && $\pm \sqrt{2} \Omega_{\rm T}$  \\
\hline
$\ket{1,1},\ket{0,3}$ && $\pm \sqrt{6} \Omega_{\rm T}$  \\
\hline
$\ket{2,0},\ket{1,2},\ket{0,4}$ && $ 0, \pm 4 \Omega_{\rm T}$ \\
\hline
$\ket{2,1},\ket{1,3},\ket{0,5}$ && $0, \pm 4 \sqrt{2} \Omega_{\rm T}$ \\
\end{tabular}
\label{tab_eigenvalues}
\end{table}

\begin{figure}[bt]
\includegraphics[width=\columnwidth]{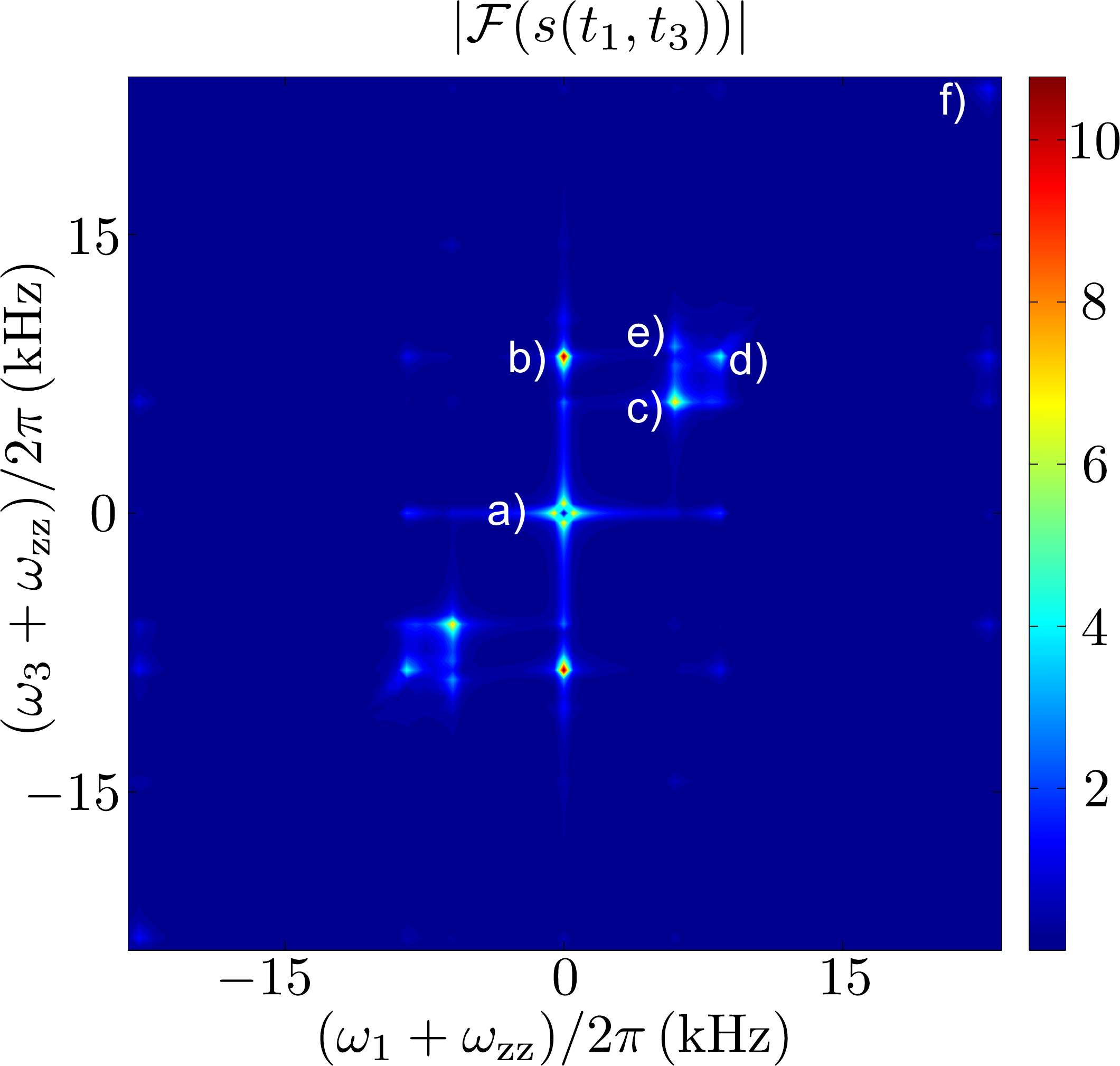}
\caption{Two-dimensional spectrum obtained for a time evolution given in~\eq{full_time_ev_supp_mat} where the free evolution is governed by $H_{\rm res}^{(3)}$,~\eq{eff_third_order_ham_app}.
The time evolution includes heating of the modes which leads to broadening of the peaks along the frequency axes. The
frequency coordinates of the points a)-f) are given in the text and allow for an identification of all appearing peaks.
The complete simulation parameters are given in the main text.}
\label{fig_third_order_supp_mat}
\end{figure}

The time evolution of the full experimental sequence for the obtention of the 2D spectrum is given
in~\eq{full_time_ev_supp_mat} with the free evolution governed by $H^{(3)}_{\rm res}$.
An experimental cycle is completed by a measurement of the zigzag mode population. In Fig.~\ref{fig_third_order_supp_mat} we show the spectrum obtained by
the aforementioned time evolution starting in a thermal state with mean phonon numbers $\bar{n}_{\rm zz} =0.7$ and $\bar{n}_{\rm str} =0.2$ for the zigzag
and stretch mode, respectively. The time evolution includes heating of the modes with heating rates $\dot{\bar{n}}_{\rm zz/str} = 0.2/0.1{\rm quanta}\cdot{\rm ms}^{-1}$.
The Hilbert spaces were truncated at $n_{\rm str,max}=6$ and $n_{\rm zz,max}=9$. The remaining parameters used in the simulations can be found in 
the main text. Note that we have substracted the bright maximum at $(-\omega_{\rm zz},-\omega_{\rm zz})$ in the center of the spectrum in order to enhance the contrast of the figure.

\begin{figure}[bt]
\includegraphics[width=\columnwidth]{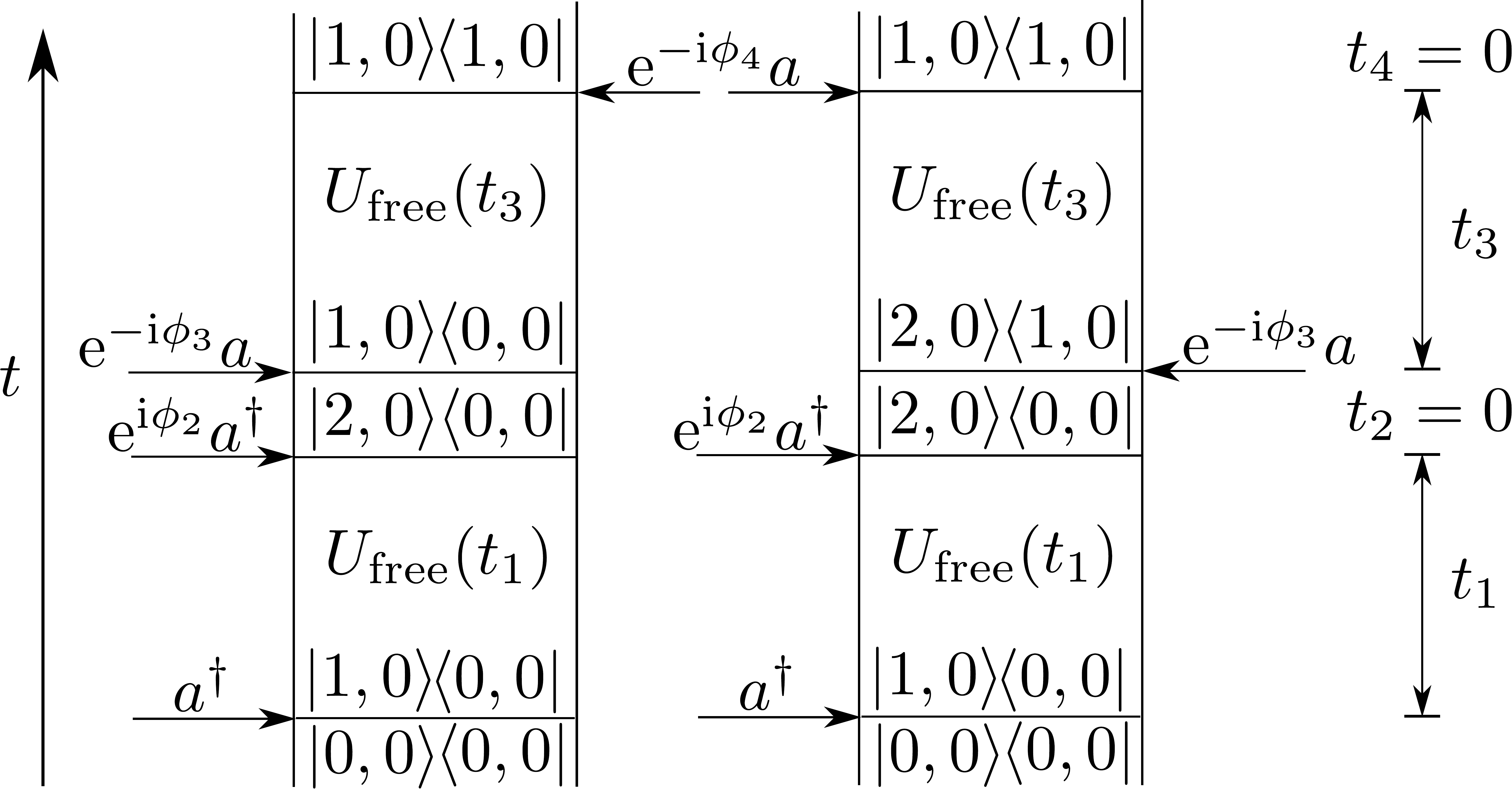}
\caption{Coherence transfer pathways leading to the dominant peaks in the spectrum shown in
Fig.~\ref{fig_third_order_supp_mat}. The pathway on the left yields a peak
at $(-\omega_{ \rm zz},-\omega_{ \rm zz})$ while the right pathway leads to peaks at $(-\omega_{ \rm zz},-\omega_{ \rm zz}\pm \sqrt{2} \Omega_{\rm T})$}
\label{fig_third_order_pathways}
\end{figure}

We will now identify the peaks appearing in the spectrum. We start by noting that the peaks in the spectrum can be
related by reflections with respect to the origin. Therefore, we will
only identify the peaks a)-f) marked in the figure, which is enough to infer
the coordinates of all other peaks. In Fig.~\ref{fig_third_order_pathways} we illustrate the two pathways leading to the
dominant peaks in
the spectrum located at a) and b) (and its symmetric counterpart). Both pathways lead to contributions which oscillate
at frequency $-\omega_{\rm zz}$ during the free evolution period $t_1$. During the second
free-evolution period, the contribution from the left pathway also oscillates with $-\omega_{\rm zz}$ while the
right pathway has time dependences $-\omega_{\rm zz}\pm \sqrt{2}\Omega_{\rm T}$.
The two frequencies in the second time evolution of the right path appear 
because the state $\ket{n_{\rm zz}=2,n_{\rm str}=0}$ may be written as a superposition of the eigenstates
corresponding to the eigenvalues $\pm \sqrt{2} \Omega_{\rm T}$ of $H^{(3)}_{\rm res}$.

The peaks identified so far correspond to the possible pathways starting from the motional ground state. In the same way, one
can find the pathways which reveal the time dependence during the free evolution periods for contributions where the
initial state contains motional excitations, leading to the understanding of the origin of the remaining spectral
peaks. The labelled peaks are then found to correspond to the coordinates:
\beq
\begin{split}
{\rm a)}:& (-\omega_{\rm zz},-\omega_{\rm zz}), \\
{\rm b)}:& (-\omega_{\rm zz},-\omega_{\rm zz}+\sqrt{2}\Omega_{\rm T}),\\
{\rm c)}:& (-\omega_{\rm zz} +(\sqrt{6}-\sqrt{2})\Omega_{\rm T},-\omega_{\rm zz} +(\sqrt{6}-\sqrt{2})\Omega_{\rm T}), \\
{\rm d)}:& (-\omega_{\rm zz}+\sqrt{2}\Omega_{\rm T},-\omega_{\rm zz}+\sqrt{2}\Omega_{\rm T}), \\
{\rm e)}:& (-\omega_{\rm zz} +(\sqrt{6}-\sqrt{2})\Omega_{\rm T}, \omega_{\rm zz} + (4-\sqrt{6})\Omega_{\rm T}),\\
{\rm f)}:& (-\omega_{\rm zz} +(\sqrt{6}+\sqrt{2})\Omega_{\rm T},-\omega_{\rm zz} +(\sqrt{6}+\sqrt{2})\Omega_{\rm T}).
\end{split}
\eeq
Accordingly, one can identify the eigenvalues of the first three manifolds in Table~\ref{tab_eigenvalues}.

\end{document}